\documentclass{aa}  

\usepackage{graphicx}
\usepackage[usenames,dvipsnames]{color}
\usepackage{pstricks}
\usepackage[percent]{overpic}
\graphicspath{{figures/}}

\usepackage{txfonts}

\usepackage{hyperref}
\hypersetup{
    bookmarks=true,         
    unicode=false,          
    pdftoolbar=true,        
    pdfmenubar=true,        
    pdffitwindow=false,     
    pdftitle={Activity of comet 67P in 2014},
    pdfauthor={C. Snodgrass et al.}, 
    pdfsubject={Planetary Science}, 
    pdfkeywords={},         
    pdfnewwindow=true,      
    colorlinks=true,        
    linkcolor=gray,         
    citecolor=blue,         
    filecolor=gray,         
    urlcolor=gray           
}

\begin{document}

   \title{Distant activity of 67P/Churyumov-Gerasimenko in 2014: Ground-based results during the Rosetta pre-landing phase\thanks{Based on observations made with ESO telescopes at the La Silla Paranal Observatory under programme IDs 592.C-0924, 093.C-0593, 094.C-0054, and at Gemini South under GS-2014B-Q-15 and GS-2014B-Q-76.}}


\titlerunning{The distant activity of 67P/Churyumov-Gerasimenko in 2014}


   \author{Colin Snodgrass
          \inst{1,2}
          \and
Emmanuel Jehin 
\inst{3}
\and
Jean Manfroid
\inst{3}
\and
Cyrielle Opitom
\inst{3}
\and
Alan Fitzsimmons
\inst{4}
\and
Gian Paolo Tozzi 
\inst{5}
\and
 Sara Faggi
\inst{5}
\and
Bin Yang
\inst{6}
\and
Matthew M. Knight
\inst{7,8}
\and
Blair C. Conn
\inst{9}
\and
Tim Lister
\inst{10}
\and
Olivier Hainaut
\inst{11}
\and
D. M. Bramich
\inst{12}
\and
Stephen~C.~Lowry
\inst{13}
\and
Agata Rozek
\inst{13}
\and
Cecilia Tubiana
\inst{2}
\and
Aur\'{e}lie Guilbert-Lepoutre
\inst{14}
  }

   \institute{
  Planetary and Space Sciences, Department of Physical Sciences, The Open University, Milton Keynes, MK7 6AA, UK         \\\email{Colin.Snodgrass@open.ac.uk}
         \and
   Max-Planck-Institute for Solar System Research, Justus-von-Liebig-Weg 3, 37077 G\"{o}ttingen, Germany
	\and
Institut d'Astrophysique et de G\'{e}ophysique, Universit\'{e} de Li\`{e}ge, Sart-Tilman, B-4000, Li\`{e}ge, Belgium
	\and
Astrophysics Research Centre, School of Mathematics and Physics, Queen's University Belfast, Belfast BT7 1NN, UK
	\and
INAF, Osservatorio Astrofisico di Arcetri, Largo E. Fermi 5, I-50 125 Firenze, Italy
	\and
European Southern Observatory, Alonso de Cordova 3107, Vitacura, Santiago, Chile
	\and
University of Maryland, College Park, MD 20742, USA
	\and
Lowell Observatory, 1400 W. Mars Hill Rd, Flagstaff, AZ 86001, USA
	\and
Gemini Observatory, Recinto AURA, Colina El Pino s/n, Casilla 603, La Serena, Chile
	\and
Las Cumbres Observatory Global Telescope Network, 6740B Cortona Drive, Goleta, CA 93117, USA
	\and
European Southern Observatory,
   Karl-Schwarzschild-Strasse 2,
   D-85748 Garching bei M\"{u}nchen, Germany
	\and
Qatar Environment and Energy Research Institute (QEERI), HBKU, Qatar Foundation, Doha, Qatar
	\and
Centre for Astrophysics and Planetary Science, School of Physical Sciences, The University of Kent, Canterbury, CT2 7NH, UK
	\and
Institut UTINAM, UMR 6213 CNRS-Universit\'{e} de Franche Comt\'{e}, Besan\c{c}on, France
           }

   \date{Received ; accepted }

 
  \abstract
   {As the ESA Rosetta mission approached, orbited, and sent a lander to comet 67P/Churyumov-Gerasimenko in 2014, a large campaign of ground-based observations also followed the comet.}
   {We constrain the total activity level of the comet by photometry and spectroscopy to place Rosetta results in context and to understand the large-scale structure of the comet's coma pre-perihelion.}
   {We performed observations using a number of telescopes, but concentrate on results from the 8m VLT and Gemini South telescopes in Chile. We use $R$-band imaging to measure the dust coma contribution to the comet's brightness and UV-visible spectroscopy to search for gas emissions, primarily using VLT/FORS. In addition we imaged the comet in near-infrared wavelengths ($JHK$) in late 2014 with Gemini-S/Flamingos 2.}
   {We find that the comet was already active in early 2014 at heliocentric distances beyond 4 au. The evolution of the total activity (measured by dust) followed previous predictions. No gas emissions were detected despite sensitive searches.}
   {The comet maintains a similar level of activity from orbit to orbit, and is in that sense predictable, meaning that Rosetta results correspond to typical behaviour for this comet. The gas production (for CN at least) is highly asymmetric with respect to perihelion, as our upper limits are below the measured production rates for similar distances post-perihelion in previous orbits.}


   \maketitle
\begin{table*}
\caption{Observations details -- FORS $R$-band imaging}
\begin{center}
\begin{tabular}{l c c c l|l c c c l}
\hline\hline
Date &  $r$ & $\Delta$ & $\alpha$ & N$_{\rm exp}$ (OK) & Date &  $r$ & $\Delta$ & $\alpha$ & N$_{\rm exp}$ (OK)\tablefootmark{a}\\
 & (au) & (au) & (deg)  &  & & (au) & (au) & (deg) \\
\hline
2014-Feb-27 & 4.39 & 4.91 & 10.4 & 10 (7)  &   2014-Aug-11 & 3.57 & 2.72 & 10.3 & 40 (15) \\
2014-Mar-12 & 4.33 & 4.68 & 11.9 & 10 (9)  &	2014-Aug-12 & 3.56 & 2.73 & 10.6 & 10 (0) \\ 
2014-Mar-13 & 4.33 & 4.66 & 12.0 & 10 (10)  &	2014-Aug-13 & 3.55 & 2.73 & 10.9 & 10 (0) \\   
2014-Mar-14 & 4.32 & 4.64 & 12.1 & 10 (10)  &	 2014-Aug-15 & 3.54 & 2.74 & 11.3 & 16 (7) \\	
2014-Apr-08 & 4.21 & 4.15 & 13.7 & 10 (0)  &	2014-Aug-16 & 3.54 & 2.74 & 11.6 & 19 (4) \\   
2014-Apr-09 & 4.20 & 4.13 & 13.8 & 10 (6)  &	2014-Aug-17 & 3.53 & 2.75 & 11.8 & 60 (17) \\  
2014-May-03 & 4.09 & 3.65 & 13.5 & 10 (10)  &	2014-Aug-18 & 3.53 & 2.75 & 12.0 & 10 (0) \\   
2014-May-06 & 4.08 & 3.60 & 13.3 & 52 (10)  &	2014-Aug-28 & 3.47 & 2.81 & 14.2 & 40 (15) \\	
2014-May-09 & 4.06 & 3.54 & 13.1 & 10 (0)  &	2014-Aug-29 & 3.46 & 2.82 & 14.4 & 10 (0) \\ 
2014-May-11 & 4.05 & 3.50 & 12.9 & 10 (10)  &	2014-Sep-11 & 3.38 & 2.92 & 16.3 & 50 (0) \\
2014-May-27 & 3.97 & 3.22 & 10.9 & 17 (0)  &	2014-Sep-18 & 3.34 & 2.98 & 17.1 & 10 (0) \\   
2014-May-31 & 3.95 & 3.16 & 10.2 & 10 (10)  &	2014-Sep-19 & 3.33 & 2.99 & 17.2 & 10 (0) \\ 
2014-Jun-04 & 3.93 & 3.10 & 9.5  & 13 (10)  &	 2014-Sep-20 & 3.33 & 3.00 & 17.3 & 50 (0) \\	
2014-Jun-05 & 3.93 & 3.09 & 9.3  & 10 (10)  &	 2014-Sep-21 & 3.32 & 3.01 & 17.3 & 7  (0) \\  
2014-Jun-08 & 3.91 & 3.04 & 8.7  & 10 (10)  &	 2014-Sep-22 & 3.32 & 3.02 & 17.4 & 22 (1) \\ 
2014-Jun-09 & 3.91 & 3.03 & 8.5  & 10 (10)  &	 2014-Sep-23 & 3.31 & 3.02 & 17.5 & 30 (0) \\ 
2014-Jun-18 & 3.86 & 2.91 & 6.3  & 10 (9)  &	 2014-Sep-24 & 3.30 & 3.03 & 17.5 & 10 (0) \\	
2014-Jun-19 & 3.86 & 2.90 & 6.1  & 10 (3)  &	 2014-Sep-26 & 3.29 & 3.05 & 17.7 & 10 (0) \\  
2014-Jun-20 & 3.85 & 2.89 & 5.9  & 10 (10)  &	 2014-Sep-27 & 3.29 & 3.06 & 17.7 & 10 (0) \\	
2014-Jun-24 & 3.83 & 2.85 & 4.9  & 35 (22)  &	 2014-Oct-09 & 3.21 & 3.16 & 18.0 & 9  (0) \\	 
2014-Jun-29 & 3.80 & 2.81 & 3.6  & 30 (25)  &	 2014-Oct-10 & 3.20 & 3.17 & 18.0 & 23 (7) \\	
2014-Jun-30 & 3.80 & 2.80 & 3.4  & 11 (10)  &	 2014-Oct-11 & 3.20 & 3.18 & 18.0 & 23 (3) \\ 
2014-Jul-01 & 3.79 & 2.79 & 3.2  & 31 (19)  &	 2014-Oct-12 & 3.19 & 3.19 & 18.0 & 10 (0) \\	 
2014-Jul-06 & 3.77 & 2.76 & 2.3  & 16 (12)  &	 2014-Oct-17 & 3.16 & 3.23 & 17.9 & 11 (4) \\	
2014-Jul-10 & 3.75 & 2.74 & 2.2  & 21 (0)  &	 2014-Oct-18 & 3.15 & 3.24 & 17.9 & 11 (3) \\ 
2014-Jul-14 & 3.72 & 2.72 & 2.8  & 11 (8)  &	 2014-Oct-22 & 3.13 & 3.27 & 17.7 & 40 (24) \\  
2014-Jul-15 & 3.72 & 2.71 & 3.0  & 10 (1)  &	 2014-Oct-23 & 3.12 & 3.28 & 17.7 & 40 (16) \\  
2014-Jul-17 & 3.71 & 2.71 & 3.4  & 10 (9)  &	 2014-Oct-24 & 3.11 & 3.29 & 17.6 & 11 (2) \\  
2014-Jul-20 & 3.69 & 2.70 & 4.2  & 18 (7)  &	 2014-Oct-25 & 3.11 & 3.29 & 17.6 & 7  (5) \\	
2014-Jul-21 & 3.68 & 2.70 & 4.5  & 20 (19)  &	 2014-Nov-14 & 2.97 & 3.42 & 15.9 & 7  (1) \\	 
2014-Jul-23 & 3.67 & 2.70 & 5.1  & 17 (12)  &	 2014-Nov-15 & 2.97 & 3.43 & 15.8 & 10 (0) \\  
2014-Jul-25 & 3.66 & 2.69 & 5.7  & 44 (18)  &	  2014-Nov-16 & 2.96 & 3.43 & 15.7 & 6  (1) \\   
2014-Jul-26 & 3.66 & 2.69 & 5.9  & 10 (0)  &	 2014-Nov-17 & 2.95 & 3.44 & 15.5 & 6  (3) \\	 
2014-Jul-27 & 3.65 & 2.69 & 6.2  & 10 (0)  &	 2014-Nov-18 & 2.95 & 3.44 & 15.4 & 28 (11) \\   
2014-Jul-28 & 3.64 & 2.69 & 6.5  & 10 (0)  &	 2014-Nov-19 & 2.94 & 3.45 & 15.3 & 13 (1) \\  
2014-Jul-30 & 3.63 & 2.70 & 7.1  & 10 (0)  &	 2014-Nov-20 & 2.93 & 3.45 & 15.2 & 9  (1) \\ 
2014-Aug-01 & 3.62 & 2.70 & 7.7  & 10 (10)  &	 2014-Nov-21 & 2.93 & 3.46 & 15.1 & 8  (0) \\  
2014-Aug-03 & 3.61 & 2.70 & 8.2  & 10 (9)  &	2014-Nov-22 & 2.92 & 3.46 & 14.9 & 10 (4) \\   
2014-Aug-06 & 3.59 & 2.71 & 9.1  & 39 (0)  &	 2014-Nov-23 & 2.91 & 3.47 & 14.8 & 7  (1) \\ 
\end{tabular}
\tablefoot{
\tablefoottext{a}{OK refers to the number of images good for photometry -- those clear within the $\rho$ = 10,000 km aperture in difference images (see text).}
}
\end{center}
\label{tab:obs-FORS} 
\end{table*}


\section{Introduction}

The ESA Rosetta spacecraft successfully arrived at, entered orbit around, and placed a lander upon comet 67P/Churyumov-Gerasimenko (hereafter 67P) in 2014. Observations from the onboard camera system OSIRIS measured the activity level of the comet between March and June, while the spacecraft was approaching but still outside of the coma \citep{Tubiana2014}. These observations showed that the comet was active, matching predictions based on previous orbits \citep{Snodgrass2013}, and displayed at least one small outburst during this time. Other remote sensing instruments detected gas in the coma (e.g. water detected by MIRO -- \citealt{Gulkis2015Sci}). Around orbit insertion in August 2014, Rosetta's suite of in situ instruments started making measurements of the coma \citep{Haessig2015Sci,Nilsson2015Sci}, while OSIRIS was able to image individual coma grains \citep{Rotundi2015Sci}. By this time, however, it was impossible to obtain images of the full extent of the coma from Rosetta's cameras, and total activity measurements were easier to make from more distant observations. 

In this paper we present results from the campaign of ground-based observations that supports the Rosetta mission, covering the Earth-based visibility period of the comet in 2014 (approximately from hibernation exit to landing). This campaign provides context observations, constraining the total activity and seeing the large-scale structure of the comet while Rosetta orbits within the very inner part of the coma. Our data comprise imaging (primarily in the $R$-band) to monitor the evolution of the coma and the total brightness of the comet and regular spectroscopic observations covering the UV and optical range, which are sensitive to the stronger comet emission bands (e.g. CN). The campaign includes observations from many telescopes\footnote{\url{http://www.rosetta-campaign.net}}, but the largest data set in 2014, while the comet was faint and in southern skies, comes from the ESO VLT. Further observations were made using Gemini South and the Las Cumbres Global Telescope Network (LCOGT) robotic telescopes. Additional results on dust photometry, using observations with the Nordic Optical Telescope (NOT), were published by \citet{Zaprudin2015}.

In section \ref{sec:obs} we describe the observations and data reduction, followed by a description of the results from photometry (total activity levels, comparison with predictions, sensitivity of ground-based observations to small outbursts) in section \ref{sec:phot}. We then describe the spectroscopic results and implications for the gas content of the coma in section \ref{sec:spec}, before comparing the ground-based results with those from Rosetta, and discussing the broader implications, in section \ref{sec:discussion}. The FORS images are also used, along with data from the OSIRIS cameras on Rosetta, to investigate the dust coma properties in more detail in a separate paper by \citet{Moreno-FORS}.


\section{Observations}\label{sec:obs}

\begin{table*}
   \caption{Spectroscopy Observations}
\begin{center}
   \begin{tabular}{l c c c c l l l c } 
\hline
Date		&  $r$	   &  $\dot{r}$	   &  $\Delta$ & $\dot{\Delta}$ & Inst. + setup & t$_{\rm exp}$	& N$_{\rm exp}$ (OK)\tablefootmark{a}   &		airmass      \\ 
		&  au	   	&   km s$^{-1}$ &  au       &   km s$^{-1}$  & 		& s		&   	   &	 min--mean--max      \\
\hline
2014-May-06      & 4.08    &  -8.40     &  3.60    &  -33.27  &  FORS 150I & 600		& 6 (6)    &	1.01--1.04-- 1.07   \\  
2014-Jun-04      & 3.93    &  -8.80     &  3.10    &  -25.78  &  FORS 150I & 600		& 6 (6)    &	1.22--1.32-- 1.45   \\  
2014-Jun-24      & 3.83    &  -9.15     &  2.85    &  -16.84  &  FORS 150I & 600		& 25 (25)  &	1.00--1.30-- 1.83   \\  
\hline	    
2014-Jul-20 - 21 & 3.69    &  -9.60     &  2.70    &  -3.35   &  FORS 300V & 900		& 29 (25)  &	1.01--1.10-- 1.35   \\  
2014-Aug-15 - 16 & 3.54    &  -10.00    &  2.74    &  8.13	 &  FORS 300V & 900		& 23 (11)  &	1.01--1.08-- 1.24   \\  
2014-Sep-23      & 3.31    &  -10.70    &  3.02    &  15.40   &  FORS 300V & 900		& 12 (4)   &	1.06--1.46-- 2.07   \\  
2014-Oct-18 - 25 & 3.13    &  -11.26    &  3.27    &  13.73   &  FORS 300V & 600, 900		& 40 (16)  &	1.27--1.61-- 2.19   \\  
2014-Nov-15 - 23 & 2.94    &  -11.74    &  3.45    &  8.34	 &  FORS 300V & 700, 750, 800, 900  & 22 (14)  &	1.75--2.27-- 2.66   \\  
\hline
2014-Nov-10 - 15 & 2.98	  & -11.63	  & 3.41	 & 9.47 	& X-SHOOTER    & 900 (UVB,NIR), 878 (VIS)      & 18 (15)       & 1.57--1.95--2.47\\
\hline
   \end{tabular}
\tablefoot{
\tablefoottext{a}{OK refers to the number of spectra used in the final analysis -- those clear of background stars where the comet signal was detected.}
}
\end{center}
   \label{tab:spec-obs}
\end{table*}

   \begin{figure}
   \centering
   \includegraphics[width=\columnwidth]{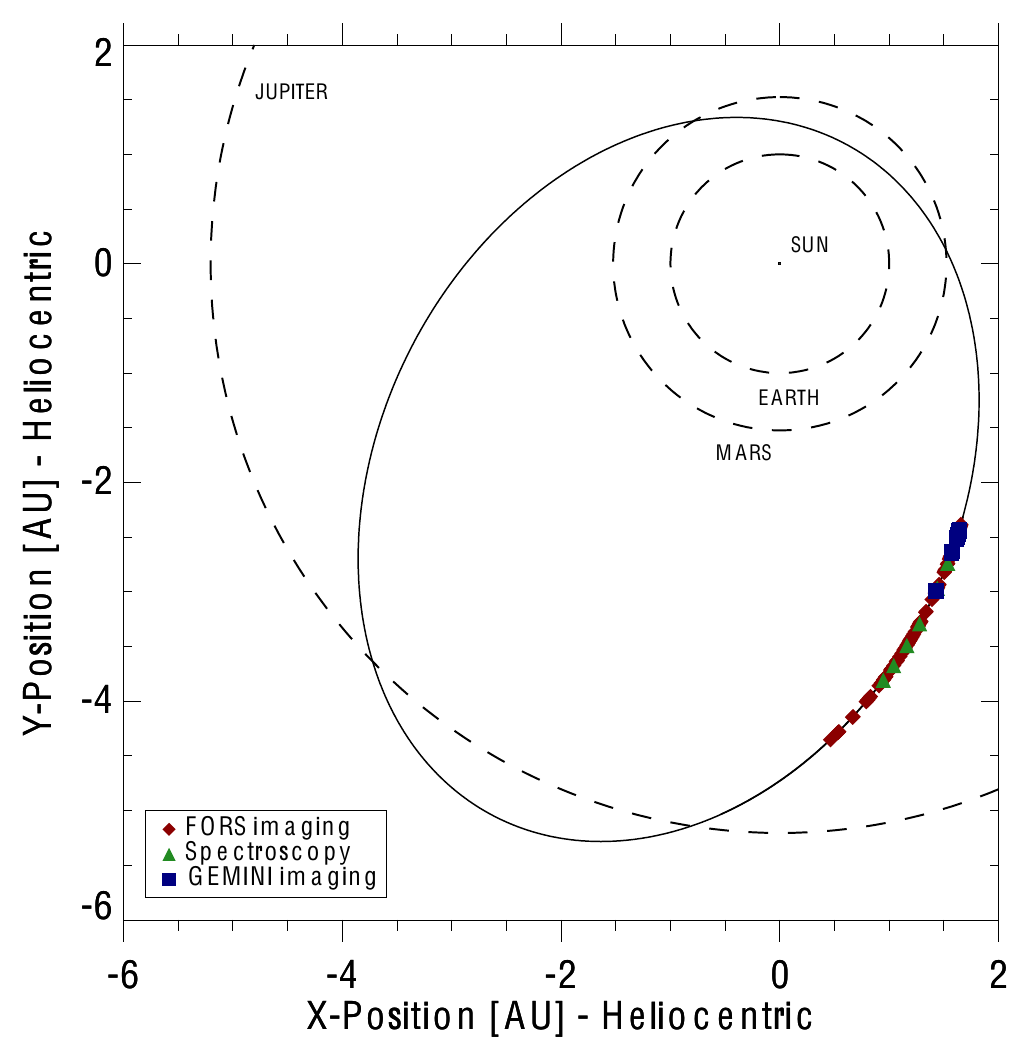}
      \caption{Orbit of 67P showing its position at the time of observations: FORS images (red), spectroscopy (green) and Gemini observations (blue).}
         \label{orbplot}
   \end{figure}

The majority of observations were acquired with the 8m ESO VLT and the FORS instrument \citep{FORS}, in both imaging and spectroscopic mode.  We also used the VLT X-SHOOTER instrument \citep{XSHOOTER} for spectroscopy around the time of the Philae landing, and the 8m Gemini South telescope with GMOS \citep{GMOS} and FLAMINGOS-2 \citep{FLAMINGOS2} instruments for imaging in $griz$ and $JHKs$ filters respectively.  The position of the comet in its orbit at the time of these observations is shown in fig.~\ref{orbplot}  -- they cover the inbound portion of the orbit between 4.4 and 2.9 au. Finally, we obtained further broadband visible wavelength images ($griz$ filters) with the LCOGT network of 1m telescopes \citep{LCOGT}. 

\subsection{VLT -- FORS}

Images were obtained using the $R$\_Special filter and in the standard 2x2 binning mode (0.25\arcsec/pixel), taken in service mode on most nights when possible, between the beginning of the visibility window in late February 2014 and the end of the window in November 2014. A log of observations is given in table \ref{tab:obs-FORS}. Images were all taken with short (50s) exposure times and sidereal tracking to allow subtraction of the crowded background star field using Difference Image Analysis (DIA -- see section \ref{sec:phot}).

Spectroscopic observations began with relatively shallow spectra (1 hour integration time) in service mode in April 2014, where the intention was to measure only the continuum slope (dust colour) and be sensitive to much higher-than-expected gas flux,  as such a discovery would have represented critical information for Rosetta mission planning. Grism 150I (central wavelength $\lambda_c = 720$ nm) was used, which gives a broad wavelength coverage (330 -- 1100 nm) and low resolution (R = 260). From June onwards deeper spectroscopic investigations were performed in visitor mode. Grism 300V ($\lambda_c = 590$ nm) was used to gain slightly higher resolution (R = 440), and therefore improved S/N in emission bands, when it was clear that the comet was bright enough to get a strong detection of the continuum. The observation geometry for each run is given in table~\ref{tab:spec-obs}. The slit width for all comet spectra was 1.31\arcsec. The slit was aligned with the parallactic angle. Solar analogues were also observed during each run, although subtle differences between the different stars observed were noted during the data reduction. To have a uniform data set, the final reduction used average spectra of the solar analogue star HD148642 for the grism 150I data and of asteroid (5) Astraea for the higher resolution data.

\subsection{VLT -- X-SHOOTER}

Around the time of the Philae landing FORS was unavailable (due to VLT UT1 M1 recoating) and our spectroscopy programme utilised X-SHOOTER on UT3 instead. X-SHOOTER has the advantage of covering a wider wavelength range, with simultaneous 300-2500 nm coverage in three arms (UVB, VIS, NIR) fed by dichroics, but has a higher resolution than was necessary for our observations and only a short slit (11\arcsec{} long). The slit widths used were 1.3\arcsec{} in the UVB arm and 1.2\arcsec{} for the VIS and NIR, i.e. the maximum widths available. Although in theory the wavelength range has the bonus of being sensitive to OH emission at 308 nm, and therefore more direct access to the water production rate, the high airmass and relative faintness of the comet at the time of observation made useful observations in the bluest order impossible. The asteroid (5) Astraea was observed as a solar analogue.

\begin{table}
\caption{Observations details -- Gemini imaging}
\begin{center}
\begin{tabular}{l c c c l}
\hline
UT Date &  $r$ & $\Delta$ & $\alpha$ &  filt. \\
 & (au) & (au) & (deg)  \\
\hline
GMOS\\
2014-Sep-20 & 3.33 & 3.00 & 17.3 &  grz \\
2014-Oct-28 & 3.09 & 3.32 & 17.4 &  grz \\
2014-Nov-10 & 3.00 & 3.40 & 16.3 &  r \\
2014-Nov-11 & 2.99 & 3.41 & 16.2 &  r \\
2014-Nov-12 & 2.99 & 3.41 & 16.1 &  r \\
2014-Nov-14 & 2.97 & 3.42 & 15.9 &  r \\
2014-Nov-17 & 2.95 & 3.44 & 15.5 &  r \\
2014-Nov-18 & 2.95 & 3.44 & 15.4 &  iz \\
\hline
\multicolumn{2}{l}{Flamingos 2}\\
2014-Sep-20 & 3.33 & 3.00 & 17.3 &  JHK \\
2014-Oct-29 & 3.08 & 3.32 & 17.3 &  JHK \\
2014-Nov-14 & 2.97 & 3.42 & 15.9 &  K \\
2014-Nov-18 & 2.95 & 3.44 & 15.4 &  J \\
\hline
\end{tabular}
\end{center}
\label{tab:obs-Gemini}
\end{table}%

\subsection{Gemini -- GMOS}

GMOS imaging in SDSS $griz$ filters was scheduled from August to November 2014. Images in $r$ were useful in complementing the FORS imaging sequence around landing time (especially during the period that FORS was not available), but only limited colour information was obtained due to bad luck in the comet falling too close to brighter stars on many of the nights when other filters were obtained. DIA techniques were not possible in these cases as the observations were taken late in the season and background images were mostly not obtained, and in some cases the comet fell near to saturated stars, which cannot be removed with DIA in any case.

\subsection{Gemini -- FLAMINGOS-2}

Gemini South was also used for near-infrared (NIR) imaging in $JHKs$ filters, also regularly spaced through the second half of the 2014 visibility window. All NIR data taken each night and in each filter were combined to produce single stacked images of the comet, as it was still relatively faint for imaging at these wavelengths (hence the need for the 8m red-sensitive Gemini telescope). The observing geometry for all Gemini data is given in table~\ref{tab:obs-Gemini}.

\subsection{LCOGT}

Towards the end of the 2014 visibility window we began long-term monitoring programmes using robotic telescopes, primarily to establish a baseline for comparison with the 2015-6 data when the comet would be brighter. The LCOGT network has a unique capability of 24-hour / day coverage of the sky, for well placed targets, and is made up primarily of 1m telescopes deployed in groups of three. For observations of 67P in 2014 we used the southern sites in Chile, Australia and South Africa. While the visibility of the comet was extremely limited from any one site, with the network we could get short observations every 8 hours, weather permitting. 

Observations from LCOGT ran from October 2nd to 19th and from November 5th to 18th, totalling 518 exposures. Each visit performed two 120s exposures in each of the SDSS $griz$ filters. The telescopes were tracked sidereally, but the comet was not trailed significantly beyond the seeing disc (FWHM $\sim$2--3\arcsec). For some periods when observations were taken at each opportunity on the network (i.e. when the $\sim$ 8 hour cadence was achieved) the field of view of subsequent observations overlapped, raising the possibility of using DIA techniques to remove stars. However, it became apparent on inspection of the data that the signal to noise achieved on the comet was not high. The comet had to be observed in twilight and at high airmass, and consequently was too faint to perform useful photometry with the 1m telescopes, with uncertainties of at least 0.2 magnitudes in some of the more promising frames. Consequently the rest of this paper uses only the data from the 8m class telescopes.


%
%

\section{Photometry}\label{sec:phot}

\subsection{Data reduction}

\subsubsection{Visible data -- FORS}

   \begin{figure}
   \centering
   \includegraphics[width=0.32\columnwidth]{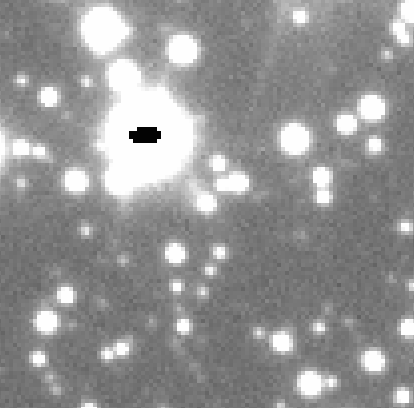}
  \includegraphics[width=0.32\columnwidth]{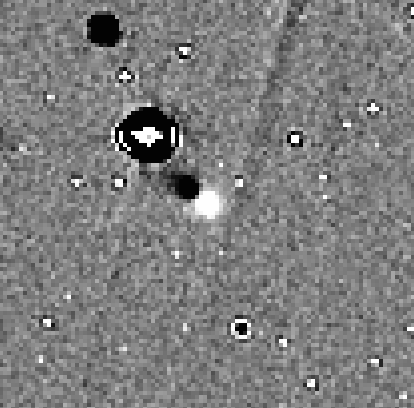}
  \includegraphics[width=0.32\columnwidth]{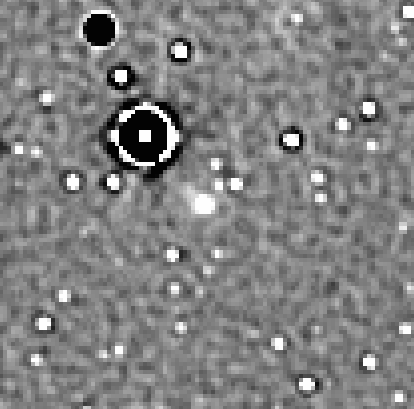}
      \caption{Example DIA processing of the same original image (left), taken on the night of Sep.~20, and showing approximately 26\arcsec, or 57,000 km, of the FORS image around the comet. The middle panel shows a DIA subtraction using the best seeing frame from the night of Sep.~20, and removes fainter stars well but demonstrates the negative imprint of the comet from its position in the template frame. The panel on the right uses a template image taken later in the year, but under poorer seeing conditions, resulting in noisier background subtraction and residuals left over from even faint stars, but without any negative comet image. Care was taken to choose the best template image for each night to maximise the number of frames where the comet was well separated from residual sources and any negative imprint.}
         \label{DIA-images}
   \end{figure}

All visible wavelength imaging observations were reduced using IDL packages from the DanIDL software suite\footnote{\url{http://www.danidl.co.uk}}, which perform basic bias subtraction and flat fielding and also allow more advanced steps, such as DIA \citep{Bramich2008,Bramich13}. DIA methods were necessary for the data on 67P taken in 2014, as they were in previous years \citep{Snodgrass2013}, due to the crowded star field that the comet was seen against. Where possible, a background template (or reference image) of the field for each night  was taken later in the year (i.e. without the comet) to provide subtraction of stars. In some cases this was not possible, especially late in the observing season, in which case one of the images taken on the same night has to be used as the background template, meaning that this template image and images close in time to it are not useful for comet photometry as the comet and its negative imprint from the template image overlap. Example images to demonstrate this are shown in fig~\ref{DIA-images}. This was complicated by the fact that DIA techniques blur the reference (background) image to match the seeing in each data frame, which meant that ideal background images had to be taken under better seeing conditions than the comet images. In many cases this was not possible. For each night the best result, in terms of maximising the number of frames in which the comet image was clear of background sources or negative imprints, was found by testing various possible background template images. 
The comet's brightness was then measured in a series of apertures, including fixed apertures in arcseconds and in physical size. The results in this paper are based on an aperture with radius $\rho=10,000$ km at the distance of the comet. Table \ref{tab:obs-FORS} reports the total number of frames taken each night and the final number used, which corresponds to the number in which there were no remaining residual stars, negative comet imprints, or cosmic rays within the 10,000 km aperture.
   
All individual images on a given night are calibrated to the photometric scale of the reference image as part of the DIA process. Absolute calibration of the photometry was performed using observations of standard stars from the \citet{Stetson} catalogue to derive zeropoints for the nights where the reference images were observed. As much of the DIA processing uses reference images obtained in only a few nights (e.g. all data obtained prior to June 24th were processed using reference images taken that night), the absolute calibration for the whole season is based on relatively few independent zeropoint calculations. There is strong internal consistency for long sequences of images based on reference images obtained on the same nights.  
In addition to this internal consistency, the derived nightly zeropoints are very stable, reflecting the high quality of Paranal and the FORS instrument, and are found to be consistent with the long term monitoring performed by ESO\footnote{\url{http://www.eso.org/observing/dfo/quality/FORS2/qc/zeropoints/zeropoints.html}}. Only one night in the final data set (October 17th) relies on a background template taken in non-photometric conditions -- this night was processed using a reference image taken the same night, and the presence of cirrus was noted by the observers. Comparing the calibrated comet brightness this night with those around it, these clouds do not appear to have had a significant effect on the results, and apparently did not affect the comet or standard star observations.

\subsubsection{NIR data -- FLAMINGOS-2}

NIR data were reduced with the IRAF packages provided by Gemini, which were then shifted and stacked based on the predicted motion of the comet.
Calibration of the NIR data was performed using 2MASS stars in the field \citep{2MASS}, measured in shorter exposures of the same field. Stars fainter than the 2MASS stars were then used to link these calibration frames with a star field frame stacked in the same way as the comet frame, but without the additional velocities added, in order to get zeropoints appropriate for the comet exposures. DIA methods were not used for the NIR data as suitable background templates were not available, but we found that these were not needed in most cases. This is due to the fact that NIR observations were taken towards the end of the season in less crowded fields, which in any case appear less crowded in the NIR, the good seeing in these images, and the fact that longer sequences of short exposures were median averaged together, removing faint stars. In the September data set the $K$-band data does appear to contain some remaining flux from a background star, resulting in an over-estimation of the brightness of the comet in that instance. Apertures of radius 5 pixels (0.9\arcsec) were used for all NIR data, which gave a good balance between collecting all the flux from the comet (the frames have a seeing with FWHM $\sim$ 2--3 pixels) while avoiding remaining background stars. The use of 2MASS stars in the field, also observed with a 5 pixel radius aperture, includes an implicit aperture correction in the calibration. The use of such aperture corrections in calibration (technically only appropriate for point sources) is justified as the comet does not appear to be visibly extended in the NIR data; in any case the corrections are much smaller than the final photometric uncertainty.

\subsection{Results}


   \begin{figure}
   \centering
   \begin{overpic}[width=0.47\columnwidth,trim=0 0 30 0, clip]{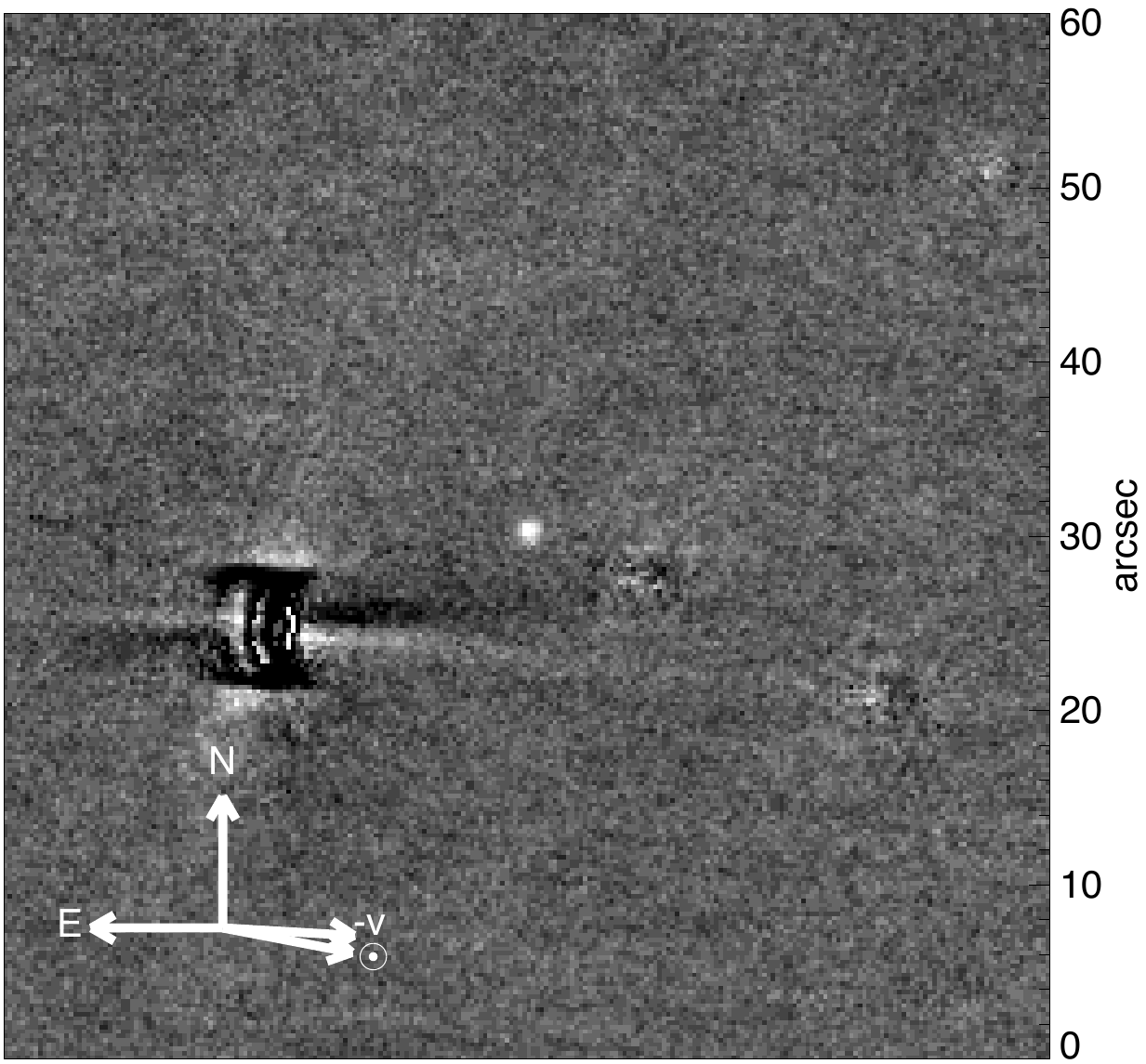}\put(5,85){\textcolor{white}{Feb}}\end{overpic}
   \begin{overpic}[width=0.514\columnwidth]{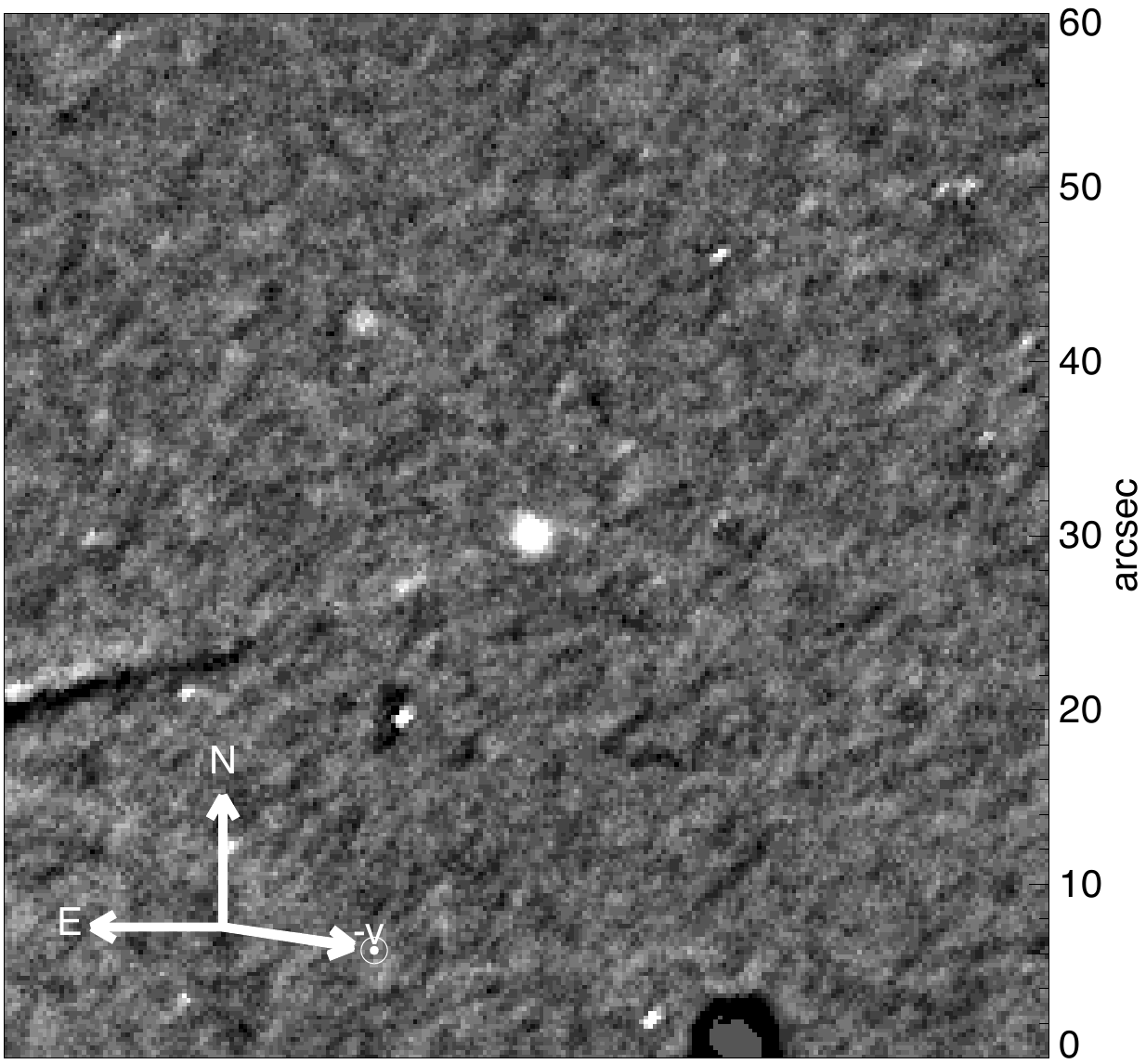}\put(5,80){\textcolor{white}{May}}\end{overpic} \\
   \begin{overpic}[width=0.47\columnwidth,trim=0 0 30 0, clip]{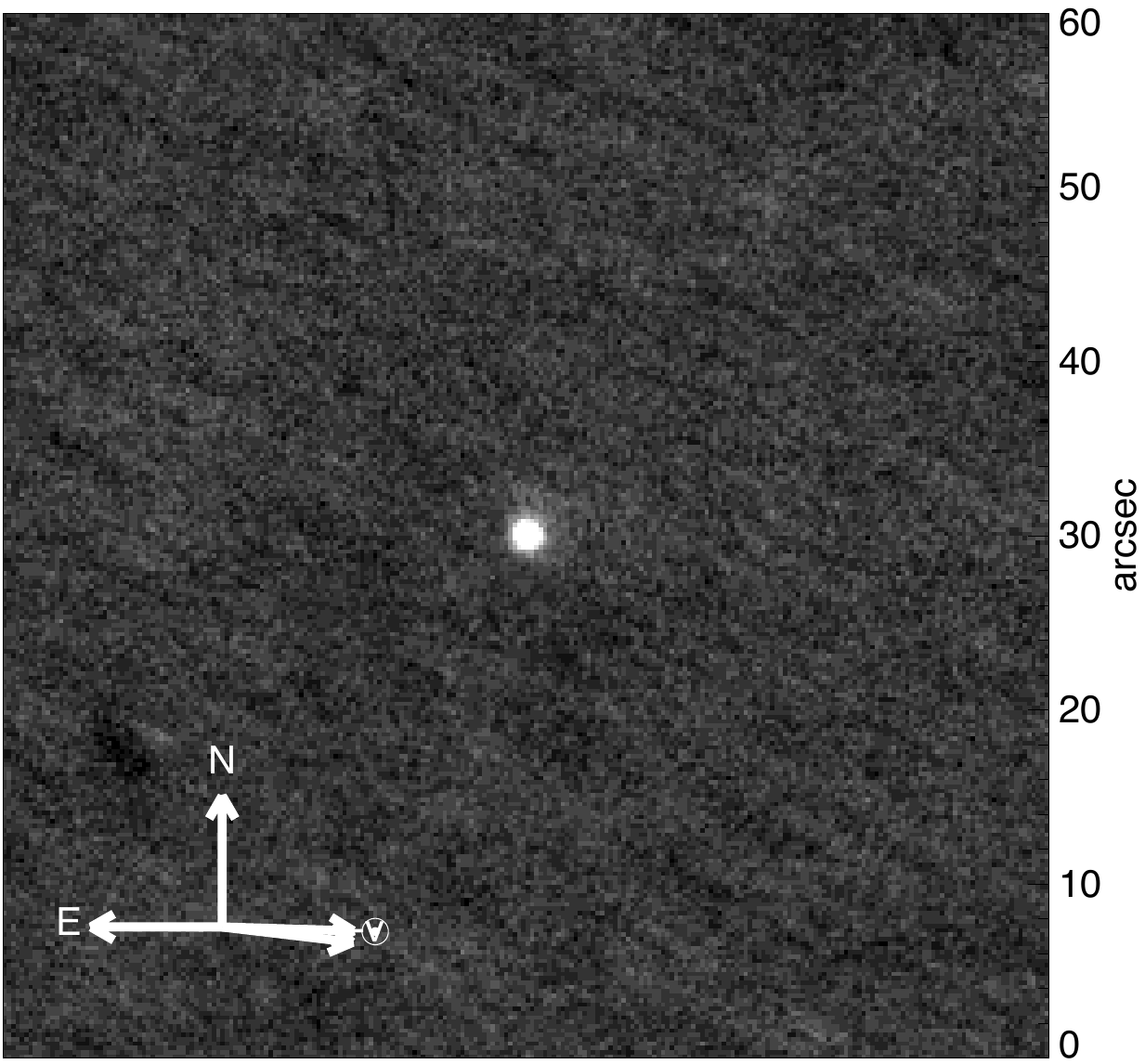}\put(5,85){\textcolor{white}{Jun}}\end{overpic}
   \begin{overpic}[width=0.514\columnwidth]{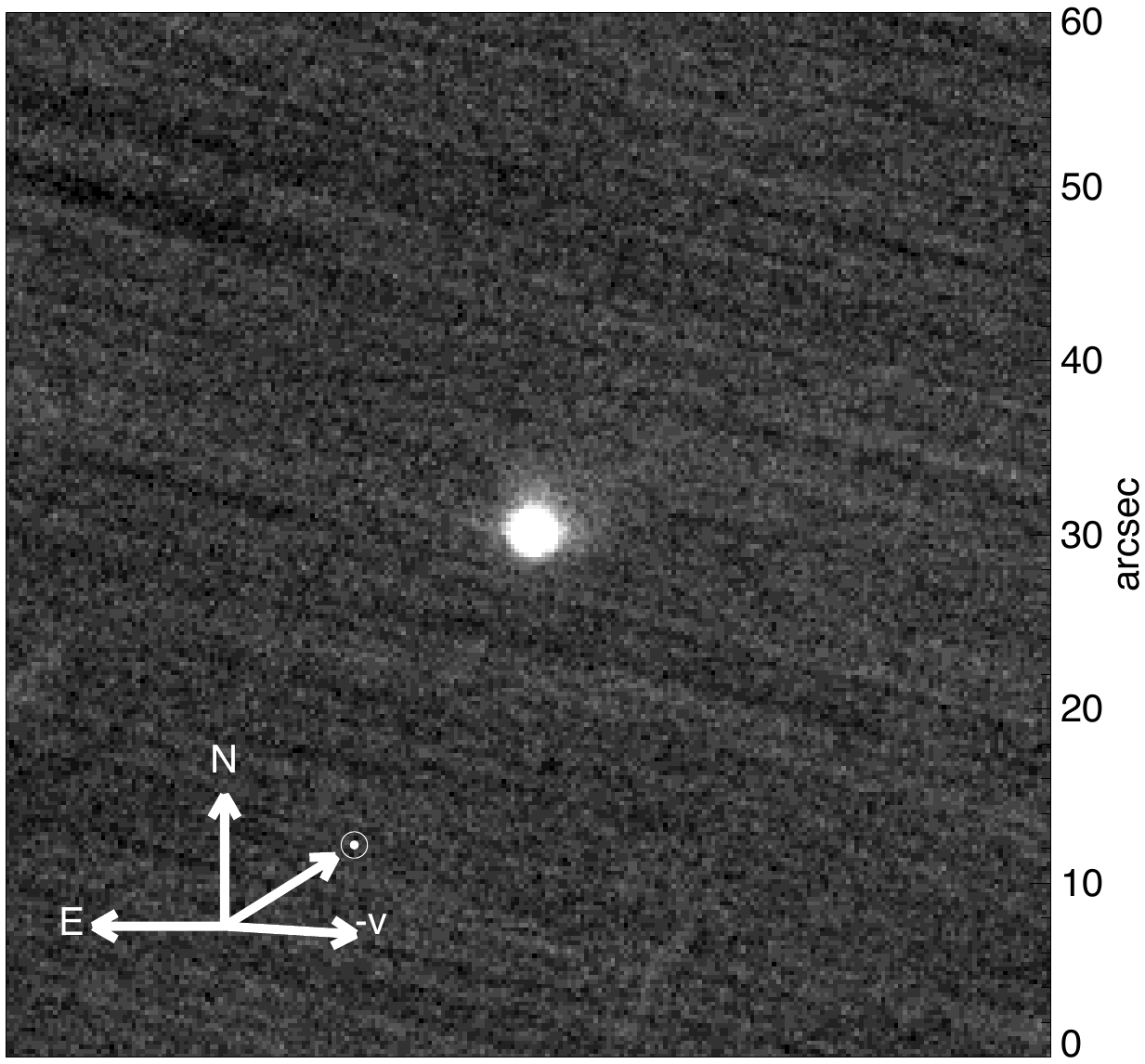}\put(5,80){\textcolor{white}{Jul}}\end{overpic}  \\
   \begin{overpic}[width=0.47\columnwidth,trim=0 0 30 0, clip]{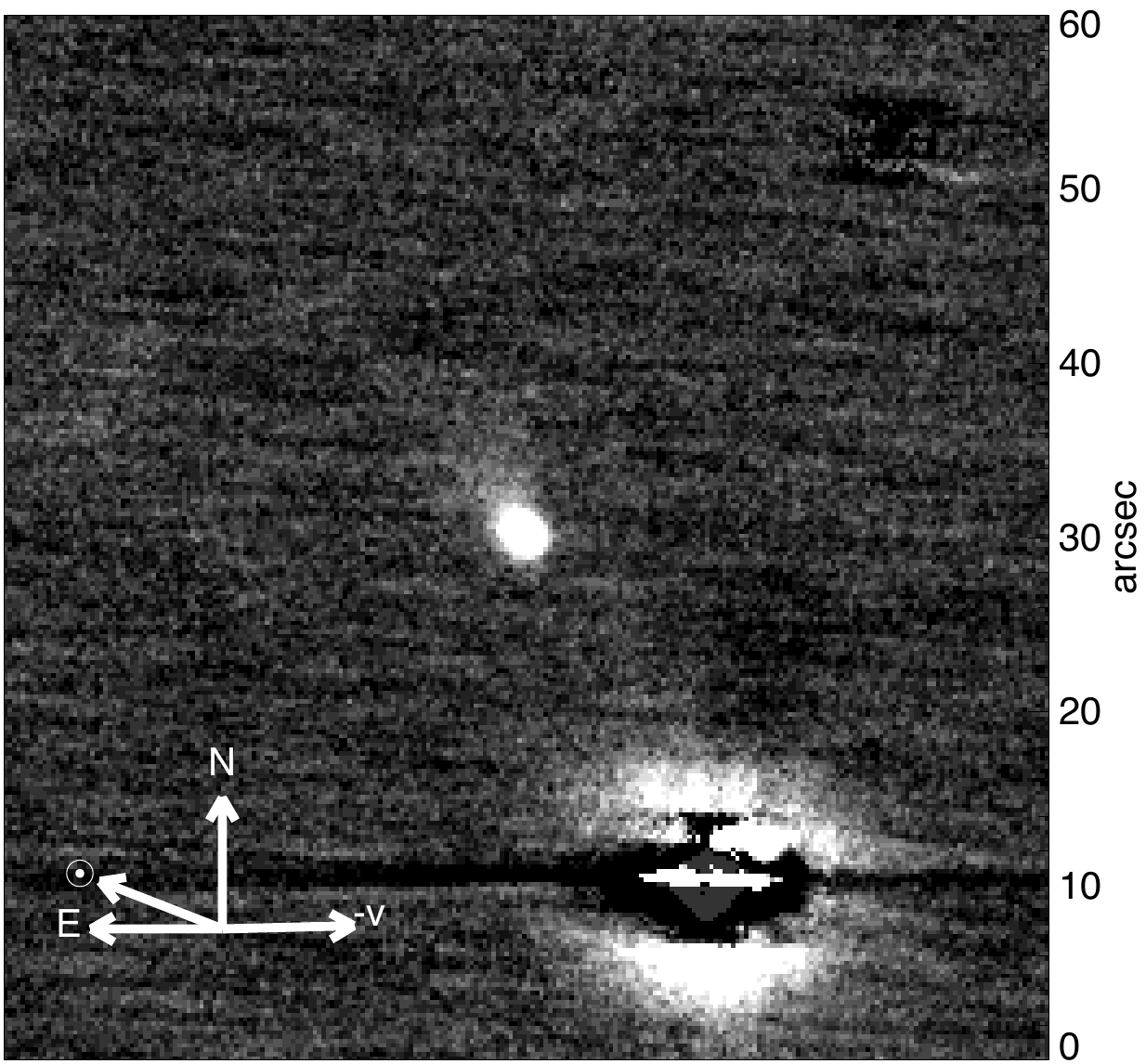} \put(5,85){\textcolor{white}{Aug}}\end{overpic}
   \begin{overpic}[width=0.51\columnwidth]{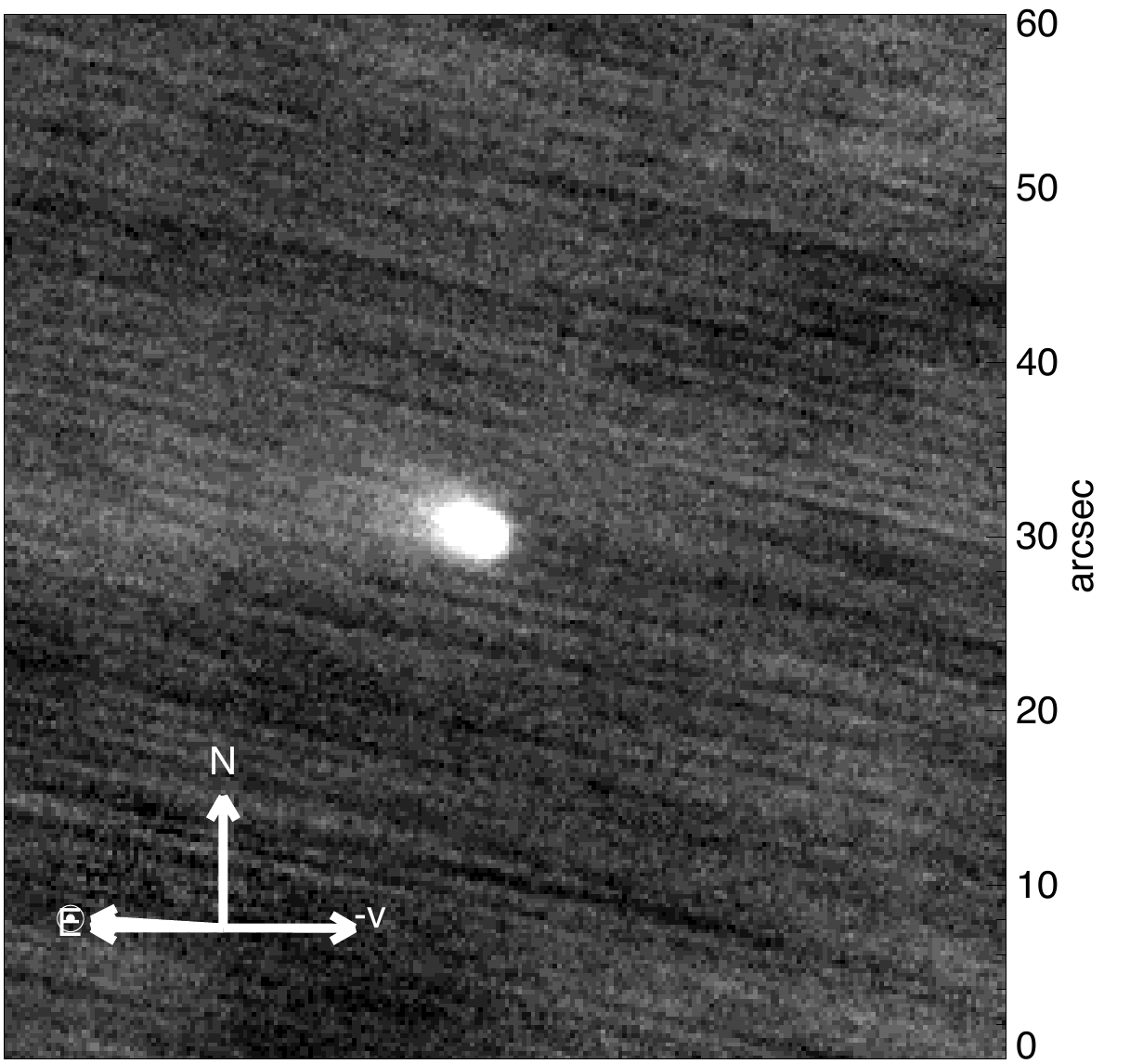}\put(5,80){\textcolor{white}{Oct}}\end{overpic}
      \caption{Images of the comet from VLT/FORS. Median combinations of every good $R$-band difference image taken on the nights of Feb.~27, May~3, Jun.~4, Jul.~1, Aug.~1 and Oct.~22. Arrows indicate the direction of the anti-velocity ($-v$) and anti-Solar ($\odot$) directions.}
         \label{morphology}
   \end{figure}

   \begin{figure}
   \centering
   \includegraphics[width=0.51\columnwidth,trim=40 137 128 135,clip]{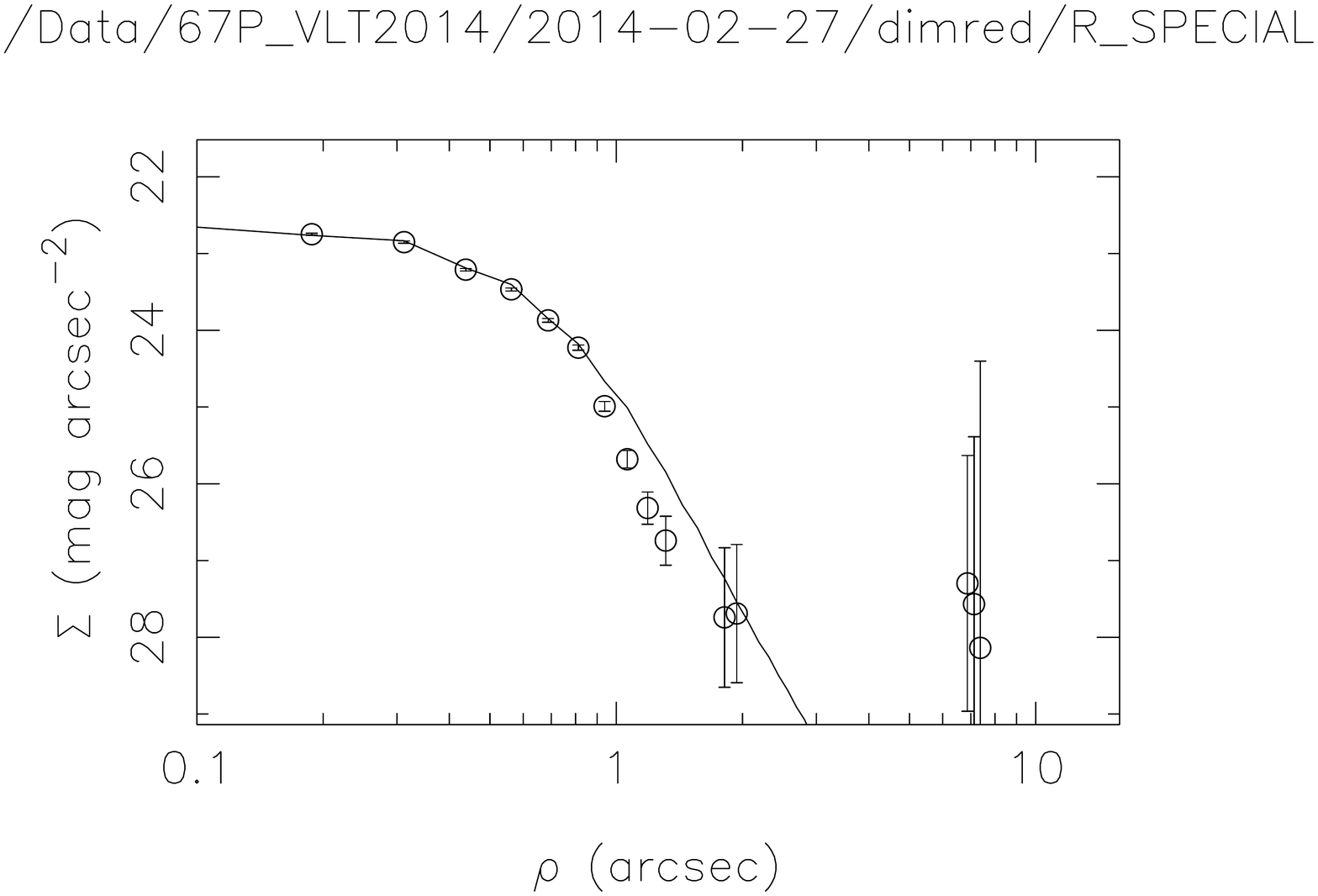}
   \includegraphics[width=0.48\columnwidth,trim=78 137 128 135,clip]{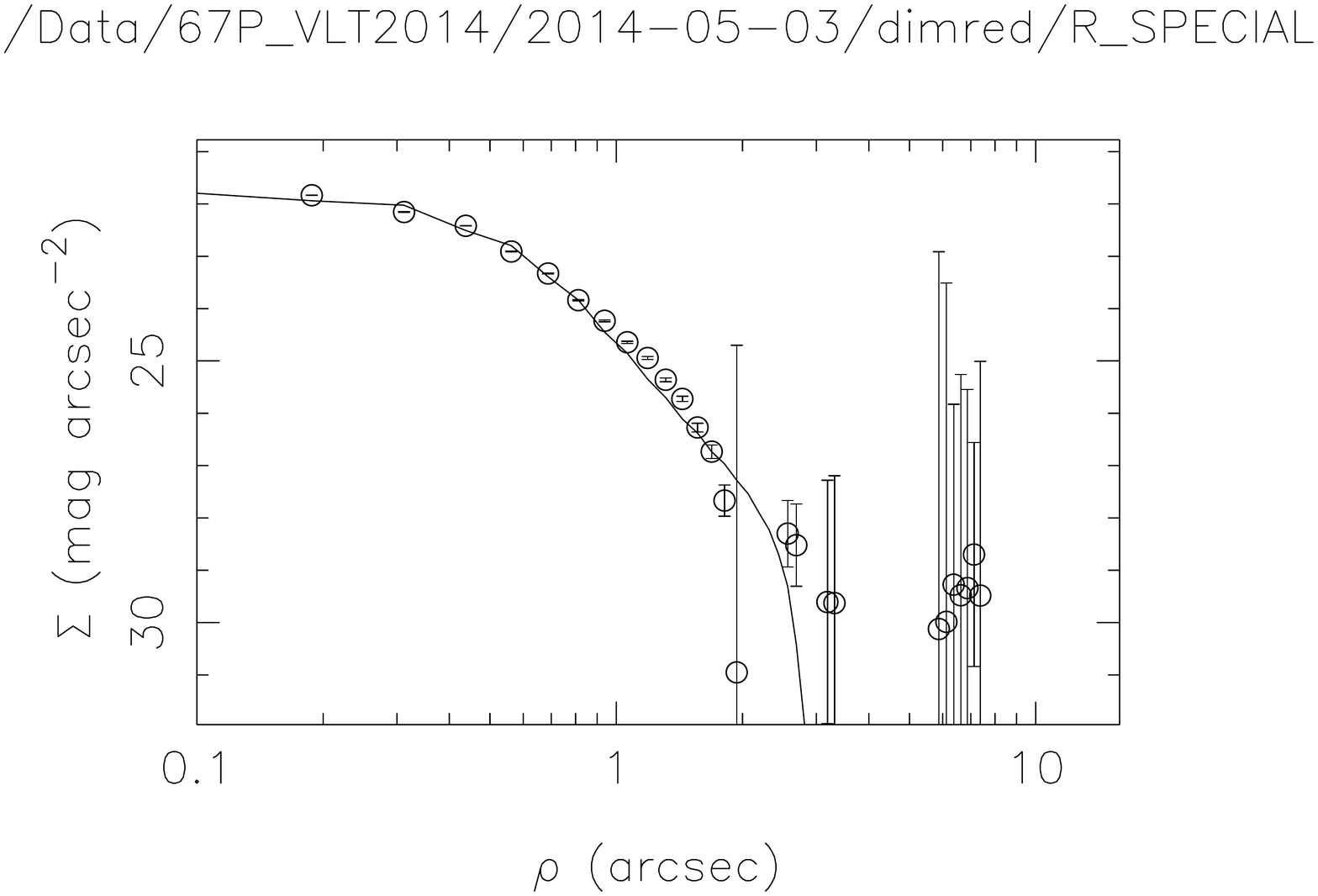} \\
   \includegraphics[width=0.51\columnwidth,trim=40 137 128 120,clip]{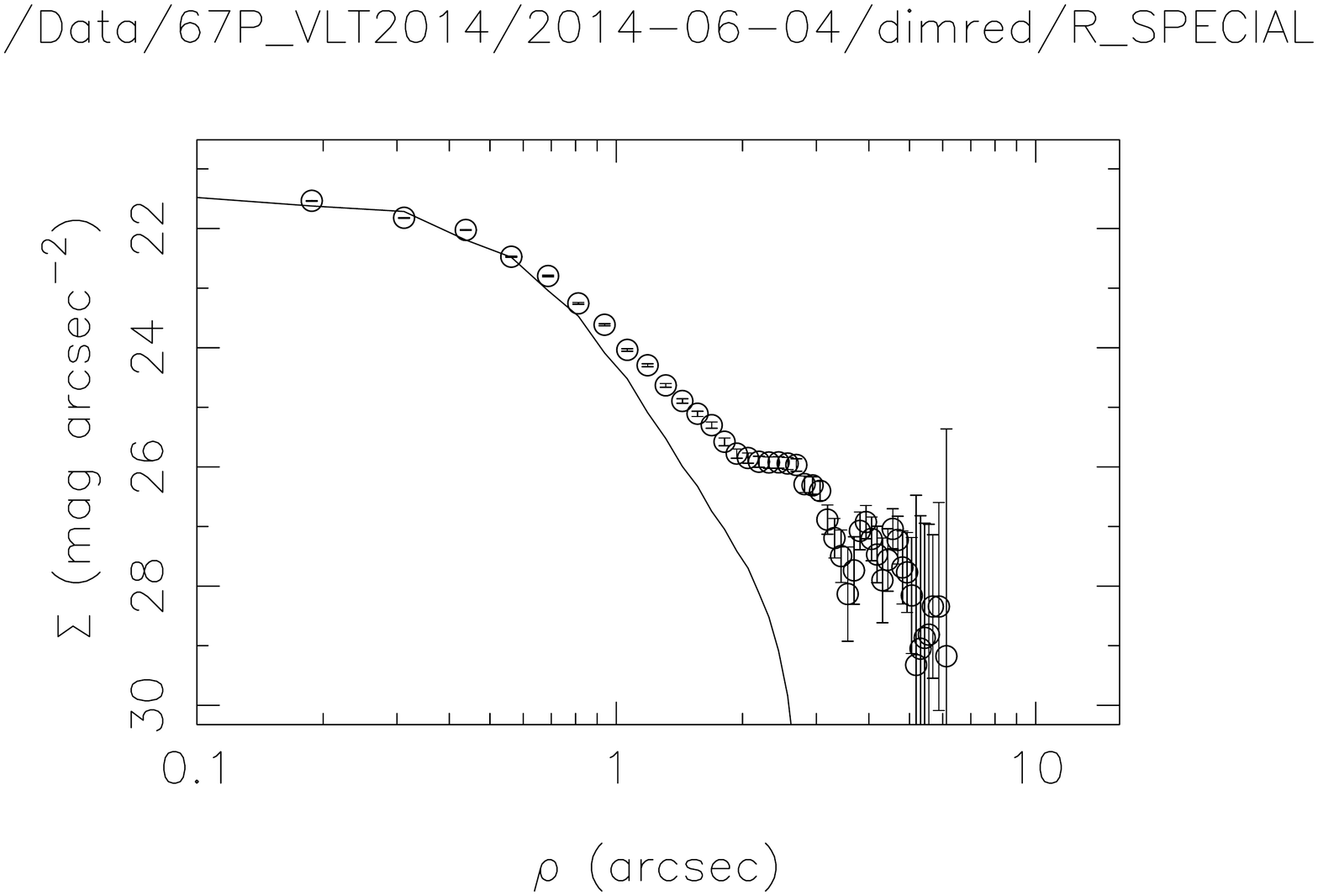}
   \includegraphics[width=0.48\columnwidth,trim=78 137 128 120,clip]{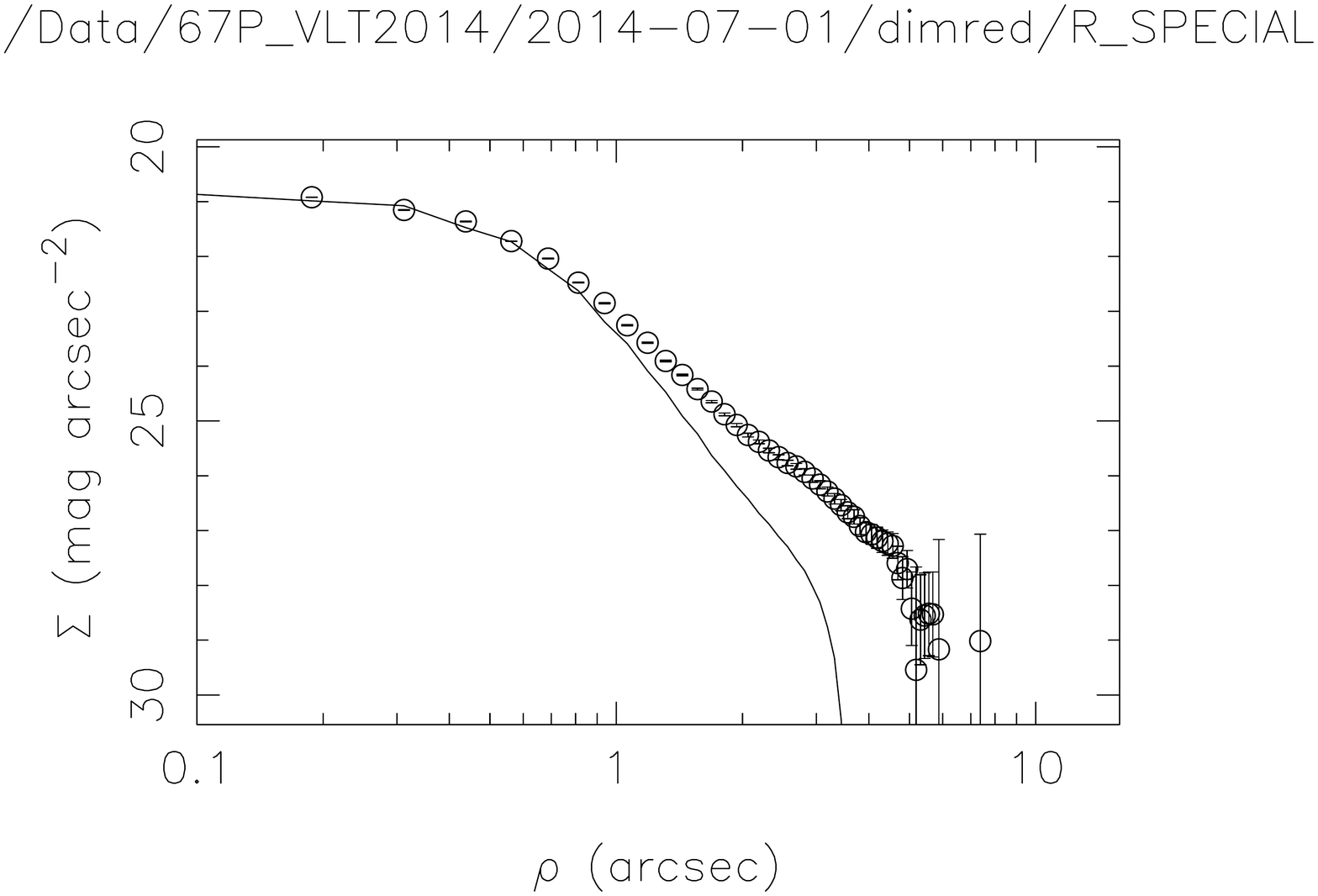}  \\
   \includegraphics[width=0.51\columnwidth,trim=40 40 128 120,clip]{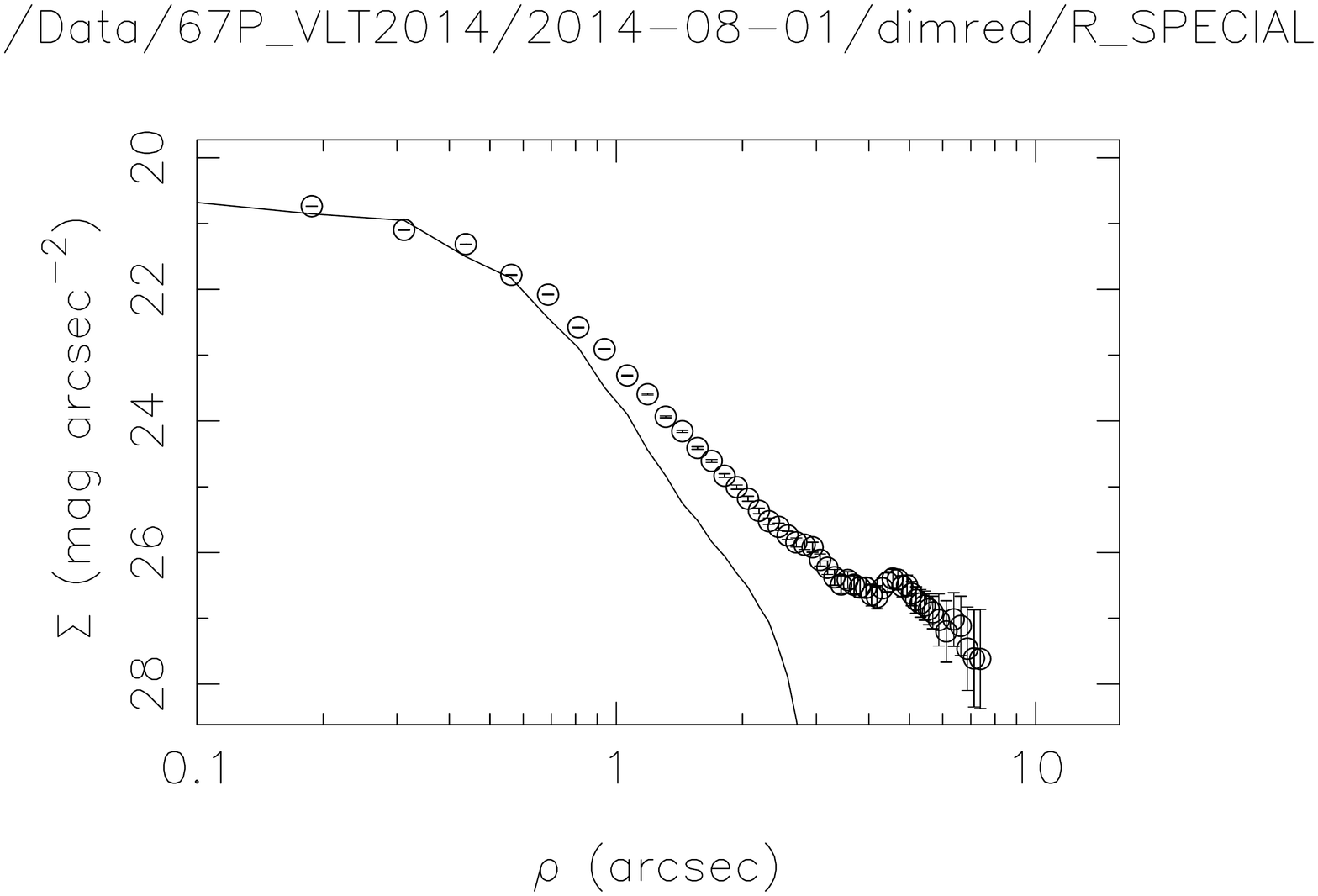} 
   \includegraphics[width=0.48\columnwidth,trim=78 40 128 120,clip]{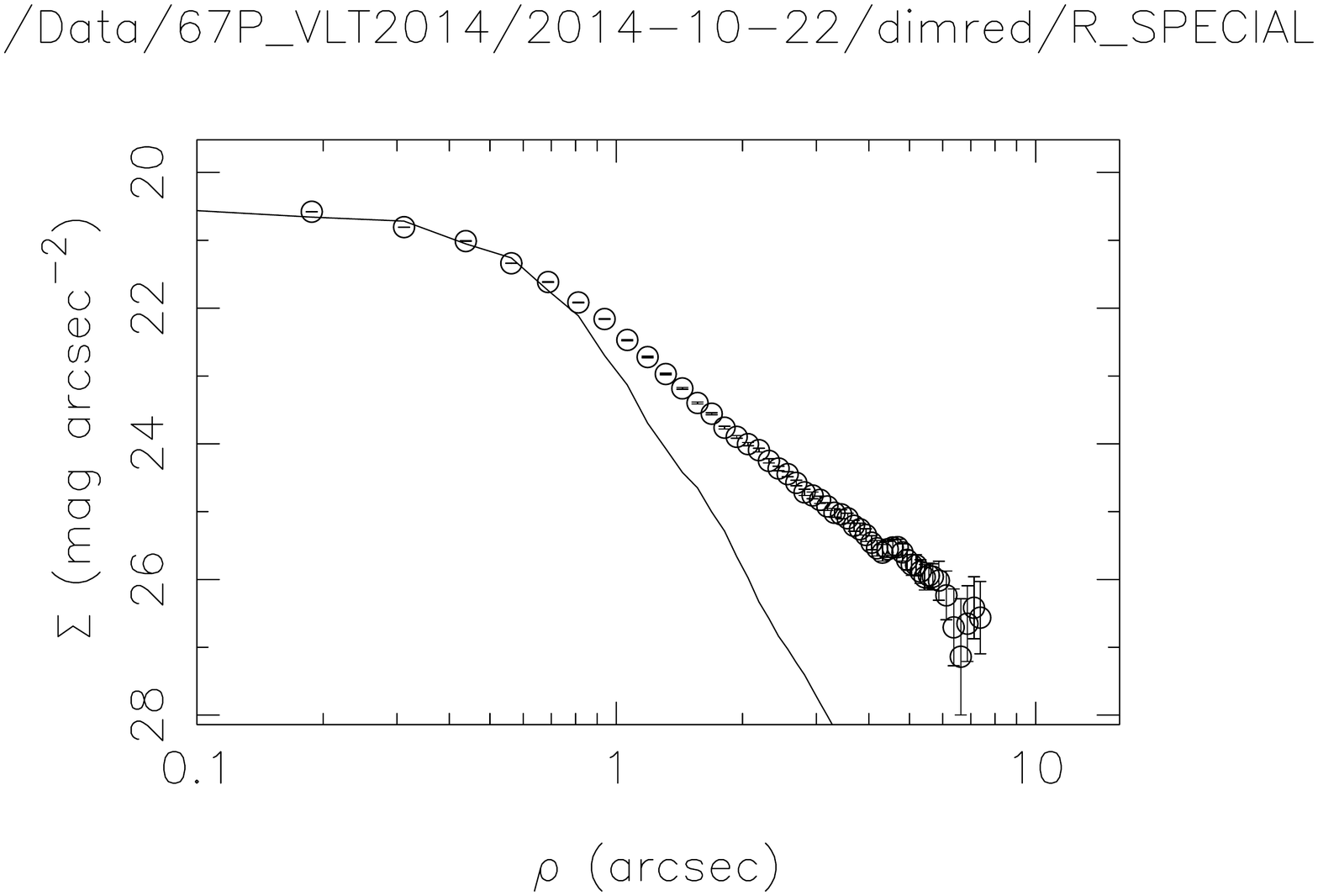}
      \caption{Surface brightness profiles for the images shown in fig.~\ref{morphology}. The points are the comet profile, the line shows the average PSF of the equivalent stacked frame, measured from stars as part of the DIA process.}
         \label{profiles}
   \end{figure}

A representative selection of images (fig.~\ref{morphology}) show the development of a visible coma during 2014. These images are median combinations of all good difference images from the night, and were selected to give relatively deep images (i.e. nights where the comet was clear of residuals from the background subtraction in a good number of frames). In the first images no coma was apparent, but photometric measurements indicated that the comet was already active as it was brighter than would be expected for a bare nucleus. Surface brightness profiles (fig.~\ref{profiles}) show that the comet appeared to be a point source in February and May, but an extended coma was detectable from June onwards. As predicted by \citet{Snodgrass2013}, resolved activity was visible from approximately July, at a heliocentric distance of $\sim$ 3.7 au. By the time that a coma was visible in ground-based images resolved activity was clearly seen in images from Rosetta, with an obvious coma visible following an outburst at the end of April \citep{Tubiana2014}. More images, from both FORS and OSIRIS, and more detailed modelling of the dust coma to explain their morphology, are presented by \citet{Moreno-FORS}. 


\begin{table}
\caption{FORS photometry}
\begin{center}
\begin{tabular}{l c c c c}
\hline
UT Date &  r & $m_R$ & $m_R$(r,1,0) & $Af\rho$ \\
 & au & \multicolumn{2}{c}{($\rho$ = 10 000 km)} & cm \\
\hline
2014-Feb-27 & 4.39 & 22.53 $\pm$ 0.11 & 18.87 & 7.1   \\ 
2014-Mar-12 & 4.33 & 22.38 $\pm$ 0.07 & 18.80 & 7.4   \\ 
2014-Mar-13 & 4.33 & 22.26 $\pm$ 0.05 & 18.67 & 8.3   \\ 
2014-Mar-14 & 4.32 & 21.88 $\pm$ 0.08 & 18.30 & 11.6  \\ 
2014-Apr-09 & 4.20 & 22.20 $\pm$ 0.07 & 18.85 & 6.7   \\ 
2014-May-03 & 4.09 & 21.55 $\pm$ 0.03 & 18.47 & 9.0   \\ 
2014-May-06 & 4.08 & 21.27 $\pm$ 0.03 & 18.23 & 11.1  \\ 
2014-May-11 & 4.05 & 21.27 $\pm$ 0.03 & 18.29 & 10.4  \\ 
2014-May-31 & 3.95 & 20.92 $\pm$ 0.03 & 18.22 & 10.5  \\ 
2014-Jun-04 & 3.93 & 20.91 $\pm$ 0.03 & 18.26 & 10.0  \\ 
2014-Jun-05 & 3.93 & 20.91 $\pm$ 0.03 & 18.28 & 9.8   \\ 
2014-Jun-08 & 3.91 & 20.71 $\pm$ 0.03 & 18.13 & 11.2  \\ 
2014-Jun-09 & 3.91 & 20.82 $\pm$ 0.04 & 18.25 & 10.0  \\ 
2014-Jun-18 & 3.86 & 20.52 $\pm$ 0.04 & 18.07 & 11.5  \\ 
2014-Jun-19 & 3.86 & 20.40 $\pm$ 0.05 & 17.96 & 12.7  \\ 
2014-Jun-20 & 3.85 & 20.60 $\pm$ 0.03 & 18.17 & 10.4  \\ 
2014-Jun-24 & 3.83 & 20.44 $\pm$ 0.04 & 18.07 & 11.3  \\ 
2014-Jun-29 & 3.80 & 20.20 $\pm$ 0.02 & 17.88 & 13.3  \\ 
2014-Jun-30 & 3.80 & 20.09 $\pm$ 0.03 & 17.79 & 14.5  \\ 
2014-Jul-01 & 3.79 & 20.22 $\pm$ 0.02 & 17.92 & 12.7  \\ 
2014-Jul-06 & 3.77 & 19.91 $\pm$ 0.03 & 17.66 & 16.1  \\ 
2014-Jul-14 & 3.72 & 20.06 $\pm$ 0.05 & 17.84 & 13.3  \\ 
2014-Jul-15 & 3.72 & 20.09 $\pm$ 0.21 & 17.86 & 13.0  \\ 
2014-Jul-17 & 3.71 & 20.07 $\pm$ 0.06 & 17.84 & 13.1  \\ 
2014-Jul-20 & 3.69 & 20.05 $\pm$ 0.07 & 17.80 & 13.4  \\ 
2014-Jul-21 & 3.68 & 20.09 $\pm$ 0.04 & 17.84 & 12.9  \\ 
2014-Jul-23 & 3.67 & 20.15 $\pm$ 0.05 & 17.89 & 12.3  \\ 
2014-Jul-25 & 3.66 & 20.09 $\pm$ 0.04 & 17.82 & 13.0  \\ 
2014-Aug-01 & 3.62 & 20.21 $\pm$ 0.06 & 17.90 & 11.8  \\ 
2014-Aug-03 & 3.61 & 20.12 $\pm$ 0.06 & 17.80 & 12.9  \\ 
2014-Aug-11 & 3.57 & 20.21 $\pm$ 0.05 & 17.82 & 12.3  \\ 
2014-Aug-15 & 3.54 & 20.12 $\pm$ 0.08 & 17.70 & 13.6  \\ 
2014-Aug-16 & 3.54 & 19.90 $\pm$ 0.03 & 17.48 & 16.7  \\ 
2014-Aug-17 & 3.53 & 20.00 $\pm$ 0.03 & 17.56 & 15.4  \\ 
2014-Aug-28 & 3.47 & 20.39 $\pm$ 0.03 & 17.86 & 11.3  \\ 
2014-Sep-22 & 3.32 & 20.12 $\pm$ 0.20 & 17.37 & 16.1  \\ 
2014-Oct-10 & 3.20 & 20.09 $\pm$ 0.04 & 17.22 & 17.4  \\ 
2014-Oct-11 & 3.20 & 19.49 $\pm$ 0.07 & 16.62 & 30.1  \\ 
2014-Oct-17 & 3.16 & 19.52 $\pm$ 0.08 & 16.62 & 29.3  \\ 
2014-Oct-18 & 3.15 & 19.84 $\pm$ 0.04 & 16.93 & 22.0  \\ 
2014-Oct-22 & 3.13 & 19.57 $\pm$ 0.03 & 16.64 & 28.1  \\ 
2014-Oct-23 & 3.12 & 19.63 $\pm$ 0.02 & 16.70 & 26.7  \\ 
2014-Oct-24 & 3.11 & 19.53 $\pm$ 0.06 & 16.60 & 29.1  \\ 
2014-Oct-25 & 3.11 & 19.68 $\pm$ 0.06 & 16.74 & 25.4  \\ 
2014-Nov-14 & 2.97 & 19.36 $\pm$ 0.09 & 16.37 & 32.7  \\ 
2014-Nov-16 & 2.96 & 19.27 $\pm$ 0.09 & 16.28 & 35.2  \\ 
2014-Nov-17 & 2.95 & 19.36 $\pm$ 0.05 & 16.37 & 32.3  \\ 
2014-Nov-18 & 2.95 & 19.39 $\pm$ 0.03 & 16.39 & 31.4  \\ 
2014-Nov-19 & 2.94 & 19.27 $\pm$ 0.11 & 16.28 & 34.8  \\ 
2014-Nov-20 & 2.93 & 19.31 $\pm$ 0.10 & 16.32 & 33.3  \\ 
2014-Nov-22 & 2.92 & 19.33 $\pm$ 0.06 & 16.34 & 32.5  \\ 
2014-Nov-23 & 2.91 & 19.22 $\pm$ 0.10 & 16.23 & 35.8  \\ 
\hline
\end{tabular}
\end{center}
\label{tab:phot}
\end{table}%

   \begin{figure}
   \centering
   \includegraphics[width=\columnwidth,trim=40 35 85 80,clip]{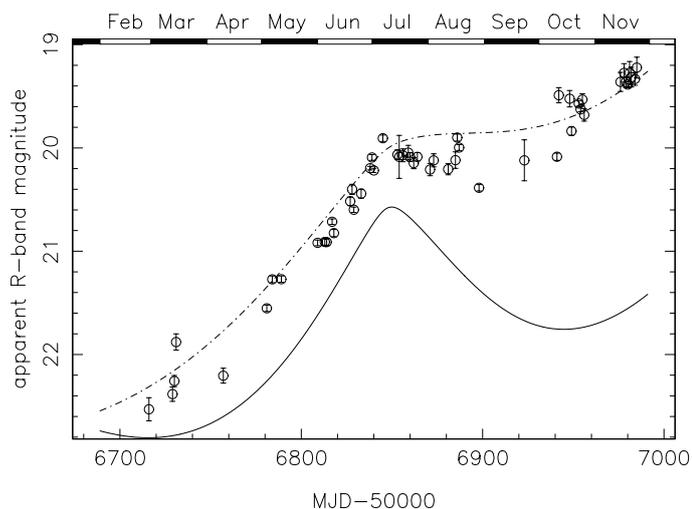}
      \caption{$R$-band magnitude, measured within an aperture of $\rho$ = 10,000 km, through 2014. The solid line shows the expected magnitude of the inactive nucleus. The dot-dashed line shows the predicted magnitude from \citet{Snodgrass2013}. Dates are in Modified Julian Days (MJD = JD - 2400000.5), with the calendar month indicated at the top of the figure for easier comparison with dates mentioned in the text.}
         \label{lightcurve}
   \end{figure}

   \begin{figure}
   \centering
   \includegraphics[width=\columnwidth,trim=40 35 100 90,clip]{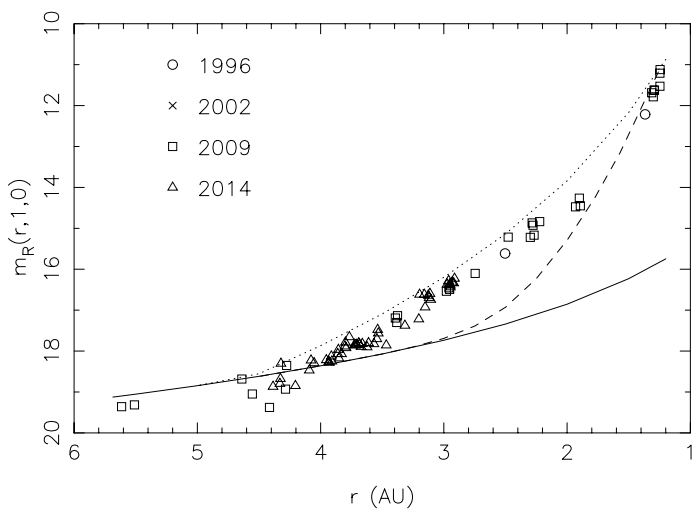}
      \caption{Heliocentric light-curve (pre-perihelion only), showing 2014 photometry compared with models (solid line = bare nucleus; dotted line = fit to previous dust photometry; dashed line = predicted brightness from expected water production) and previous data from \citet{Snodgrass2013}. See fig.~6 in that paper for further details.}
         \label{HLC}
   \end{figure}

The total $R$-band brightness of the comet, measured in an aperture equivalent in radius to $\rho=10,000$ km at the distance of the comet, is given in table~\ref{tab:phot} and shown in fig.~\ref{lightcurve}. The values are the weighted means of all measurements from each night. The light-curve also shows the predicted total brightness from \citet{Snodgrass2013}, and demonstrates that the total activity level of the comet was in good agreement with the predicted values. This is also shown in fig.~\ref{HLC}, which shows the same photometry reduced to unit geocentric distance and zero degrees phase angle (assuming a linear phase function\footnote{The phase function for comet dust is not linear, with the best description being the one given by Schleicher at \url{http://asteroid.lowell.edu/comet/dustphase.html}. However, this can be approximated by a linear phase function over the range of phase angles seen from Earth ($\alpha < 20\degr$).} with $\beta=0.02$ mag.~deg.$^{-1}$), along with the earlier data and fits from \citet{Snodgrass2013}. This implies that the activity of the comet in 2014 was very similar to that seen at the same heliocentric distances in previous orbits, and that there is no major long term trend from orbit to orbit. Preliminary photometry from the 2015 observability season suggests that the comet continues to follow this prediction towards perihelion. 2015 results will be presented in future papers.

We quantify any secular (orbit-to-orbit) change in activity level in 67P by comparing the 2009 and 2014 heliocentric magnitudes shown in fig.~\ref{HLC}. After removal of the $\propto r^{-5.2}$ dependence from the flux \citep{Snodgrass2013}, the median 2014 data is ($5\pm13$)\% fainter than the 2009 data in the same distance range ($4.4 > r > 2.9$ au). This confirms that there is no significant change in activity from orbit-to-orbit, at least for $r > 3$ au, as found for the previous three orbits \citep{Snodgrass2013}. The activity level of 67P is therefore more stable than that of the spacecraft targets 9P and 103P, which show secular decreases in activity of 20\% and 40\% respectively \citep{Schleicher2007,Meech2011b,Knight+Schleicher2013}, despite the relatively recent arrival of 67P in its current orbit \citep{Maquet2015}.

Table~\ref{tab:phot} also gives the $R$-band $Af\rho$ measurement that is commonly used to quantify dust production \citep{Ahearn84}. In this case the value has been corrected to zero phase angle, again using a linear phase function with $\beta=0.02$ mag.~deg.$^{-1}$, appropriate for cometary dust. Care must be taken in using $Af\rho$ where activity levels are low, as it assumes that any contribution to the total flux from the nucleus is negligible, but these values are consistent with previous apparitions and expectations for 2014.

The photometry indicates that there was already activity present from the earliest observations, as the points are significantly above the nucleus curve in fig.~\ref{lightcurve}. However, it is important to remember that this curve shows the average nucleus magnitude and does not include the effects of the rotational light-curve, which has a range of up to $\sim 0.5$ magnitudes. The first point, in February, should contain 78\% nucleus flux (based on the mean nucleus curve) but could be essentially all due to the nucleus (98\%) if it happened to coincide with light-curve maximum. Even without further information from Rosetta, the fact that all the points lie comfortably above the curve implies that the activity is real, as catching the rotational light-curve at maximum in every observation is statistically unlikely. We can be certain that there has to be activity already by the March observations, as here the nucleus cannot fully explain the observed brightness even at light-curve maximum (contributions of 55 - 87\% of the observed flux).

   \begin{figure}
   \centering
   \includegraphics[width=\columnwidth,trim=40 35 100 80,clip]{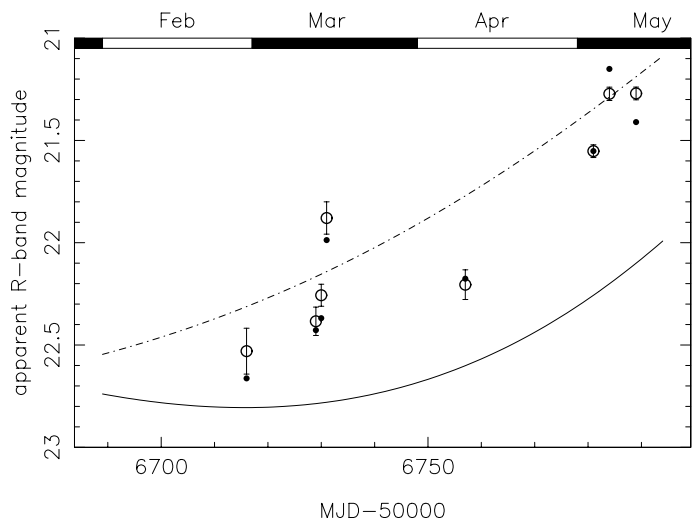}
      \caption{Zoom in on early data shown in fig.~\ref{lightcurve}, with effect of correcting points for nucleus rotation phase shown (filled circles).}
         \label{lightcurve2}
   \end{figure}

In the case of 67P, we can also be certain of the real nucleus contribution to the observed flux in early 2014, as Rosetta results give the shape \citep{Sierks2015Sci} and rotation state \citep{Mottola2014}, allowing reconstruction of the observed illuminated fraction (and therefore light-curve relative to the mean) from any viewing geometry. We generated the expected light-curve, taking the spin period from \citet{Mottola2014} as constant. This is valid for the early part of 2014 -- by the time the spin rate was observed to change significantly  \citep{Keller2015-spinup} the coma already dominates the flux. The data up until early May are shown in fig.~\ref{lightcurve2}, where the filled points have been corrected for the rotational phase at the time of observation (open circles show the original photometry). The correction is small and does not bring the photometry down to the mean nucleus curve, so there must still have been activity present (although the February point could be considered consistent with an inactive nucleus within the uncertainties).

The higher cadence of photometric monitoring in 2014 compared with previous apparitions gives a more detailed light-curve. Some structure is visible in fig.~\ref{lightcurve} beyond the overall trend identified previously. Rapid rises in March and August in particular are tempting to describe as small outbursts, although care must be taken -- the rapid variation in October, for example, is unlikely to be real despite the fact that it is larger than the formal uncertainties, and is probably due to the two fainter points (10 and 18 October) relying on direct calibration from reference images taken on the same nights, while most of the others form part of a longer sequence calibrated to reference images taken on 21 November. The variation in March cannot be explained the same way, but the August `outburst' could be. The March data (also seen in detail in fig.~\ref{lightcurve2}) shows a $\sim$ 0.5 mag.~increase in brightness over 3 days. The effect is not reduced by the nucleus corrections mentioned above, which is not surprising -- the $\sim$ 12 hour period of the nucleus means that approximately the same rotation phase is observed each night over short periods like this. If real, this event took place a few days before OSIRIS was turned on, meaning that it is quite possible that it would have been missed from Rosetta, as the outburst that OSIRIS did see at the end of April was visible for only a short time \citep{Tubiana2014}. We do not see any strong evidence for the OSIRIS outburst in the VLT photometry, but we lack observations in late April to compare with the early May data. Although this period is covered by the NOT observations by \citet{Zaprudin2015}, the S/N of those data do not allow small outbursts to be confirmed. By the time of the potential August event, Rosetta was already close enough to the nucleus that OSIRIS no longer saw the whole comet, so again small outbursts could have been missed. Careful comparison with gas pressure measurements from ROSINA, for example, may be more revealing at that time.

\begin{table}
\caption{Flamingos-2 photometry}
\begin{center}
\begin{tabular}{l c c c}
\hline
UT Date &  J & H & K \\
\hline
2014-Sep-20 &  19.63 $\pm$ 0.07 & 19.01 $\pm$ 0.07 & 18.27 $\pm$ 0.08\tablefootmark{a}\\
2014-Oct-29 &  19.18 $\pm$ 0.06 & 18.85 $\pm$ 0.06 & 18.66 $\pm$ 0.11\\
2014-Nov-14 &  -- & -- & 18.52 $\pm$ 0.14\\
2014-Nov-18 &  18.90 $\pm$ 0.06 & -- & -- \\
\hline
\end{tabular}
\tablefoot{
\tablefoottext{a}{Probably affected by background star}
}
\end{center}
\label{tab:results-Gemini}
\end{table}%

The NIR photometry from Gemini observations provides multi-colour information, although not the long timeline that is covered by the VLT observations. The resulting photometry is presented in table~\ref{tab:results-Gemini}, showing that on only one night was a good set of observations with all three NIR filters obtained. This gave colours of $(J-H) = 0.33$ and $(H-K) = 0.19$, both slightly redder than the Sun. The colour of the dust is discussed in more detail in the following section on the spectroscopic observations.

%
%

\section{Spectroscopy}\label{sec:spec}

\subsection{Data reduction}

Spectroscopic data from FORS were reduced using IRAF and MIDAS routines, and the results compared. No significant difference was found between the techniques - we report the results using IRAF routines. 
The 1-D spectrum was extracted from the debiased, flat-fielded, rectified frames, by addition over 20-30 lines. The sky background was taken in clean zones on each side of the comet spectrum.
Not all spectra were usable -- the fields were crowded and in many cases star spectra overlap the comet. Some spectra were affected by clouds and in some cases the comet was absent altogether. Spectrophotometric standards have been observed with the MOS instead of the long-slit mode in order to
have a wide (5\arcsec) slit.
Standards stars showed slight differences between the runs, hence a sensitivity curve was computed for each run.

The X-SHOOTER data were considerably more complex to deal with, due to the 3 cross dispersed arms, but the ESO X-SHOOTER pipeline \citep{XSHOOTER-pipeline} automates many of the tasks, using the the Reflex environment \citep{Reflex}. Extra care was required to process a target as faint as the comet.  
We used the Reflex pipeline to produce two dimensional order-merged and  wavelength-calibrated spectra, while the 1-D spectrum of 67P was extracted using IRAF routines. 
It was necessary to average all data taken (between the 10th and 15th of November) into a single spectrum and bin heavily in wavelength range to make use of this data.

\subsection{Results}

We calculate a number of parameters from the reduced long-slit spectra: The $(B-V)$ colour index (as a measure of spectral slope), $Af\rho$ for the dust, and (limits on) production rates for gas species. The results are summarised in table \ref{tab:spec-results}. The X-SHOOTER data were used to calculate production rate limits and to study the continuum over a longer wavelength range (into the NIR).

\begin{table*}
   \caption{Spectroscopy results -- colours and upper limits on gas production rates}
\begin{center}
   \begin{tabular}{l c c c c c c c } 
\hline
Date		&  $r$	   &   B-V		   		&  Q(C$_2$) &   Q(C$_3$) &   Q(CN)   & Af$\rho$  (BC) &  Af$\rho$  (GC) \\ 
		&  au	   &	excess\tablefootmark{a}	&	   molec. s$^{-1}$ &  molec. s$^{-1}$ &  molec. s$^{-1}$  & cm  & cm \\
\hline
2014-May-06      & 4.08    &   0.27  & < 4.0E24 & < 2.4E23 & < 3.2E24  &  12.4 $\pm$ 1.5 &  17.4 $\pm$ 2.1 \\  
2014-Jun-04      & 3.93    &   0.20  & < 2.3E24 & < 1.4E23 & < 1.9E24  &   \phantom{1}9.5 $\pm$  0.8 &  11.8 $\pm$ 1.0  \\  
2014-Jun-24      & 3.83    &   0.26  & < 1.2E24 & < 7.4E22 & < 1.0E24  &   \phantom{1}8.5 $\pm$ 0.4  &  12.2 $\pm$ 0.5  \\  
\hline	    	  
2014-Jul-20 - 21 & 3.69    &   0.16  & < 7.6E23 & < 4.4E22 & < 6.4E23	&  15.0 $\pm$ 0.8 &  19.5 $\pm$ 1.1 \\  
2014-Aug-15 - 16 & 3.54    &   0.18  & < 7.8E23 & < 4.5E22 & < 6.7E23	&  14.0 $\pm$ 1.1 &  19.1 $\pm$ 1.5 \\  
2014-Sep-23      & 3.31    &   0.09  & < 1.2E24 & < 7.5E22 & < 1.1E24	&  19.7 $\pm$ 2.8 &  26.0 $\pm$ 3.7 \\  
2014-Oct-18 - 25 & 3.13    &   0.18  & < 4.9E23 & < 2.9E22 & < 4.3E23	&  20.8 $\pm$ 1.3 &  26.1 $\pm$ 1.6 \\  
2014-Nov-15 - 23 & 2.94    &   0.17  & < 5.2E23 & < 3.1E22 & < 4.6E23	&  23.6 $\pm$ 1.6 &  30.5 $\pm$ 2.1 \\  
\hline		  
2014-Nov-10 - 15 & 2.98	   &   --    & < 2.7E23 & < 1.7E22 & < 2.1E23   &  --  &  --  \\
\hline
   \end{tabular}
\tablefoot{
\tablefoottext{a}{ $(B-V)$ excess relative to the Sun. The RMS uncertainty on each measurement is 0.07, mostly due to the calibration.}
}
\end{center}
   \label{tab:spec-results}
\end{table*}

\begin{figure}
   \centering
   \includegraphics[width=\columnwidth]{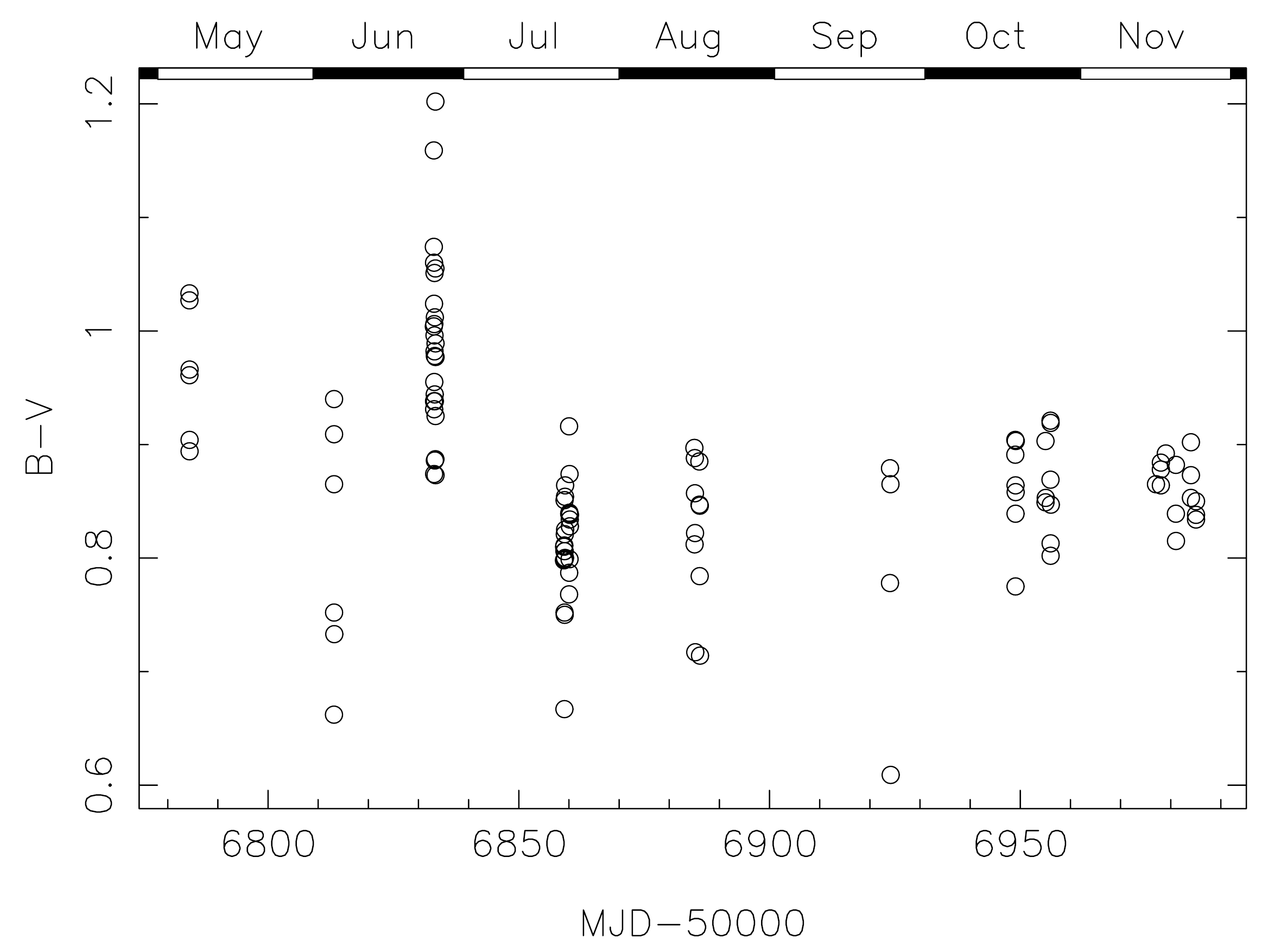}
      \caption{B-V index for all FORS spectra.}
         \label{BV-FORS-spec}
   \end{figure}

\begin{figure}
   \centering
   \includegraphics[width=\columnwidth]{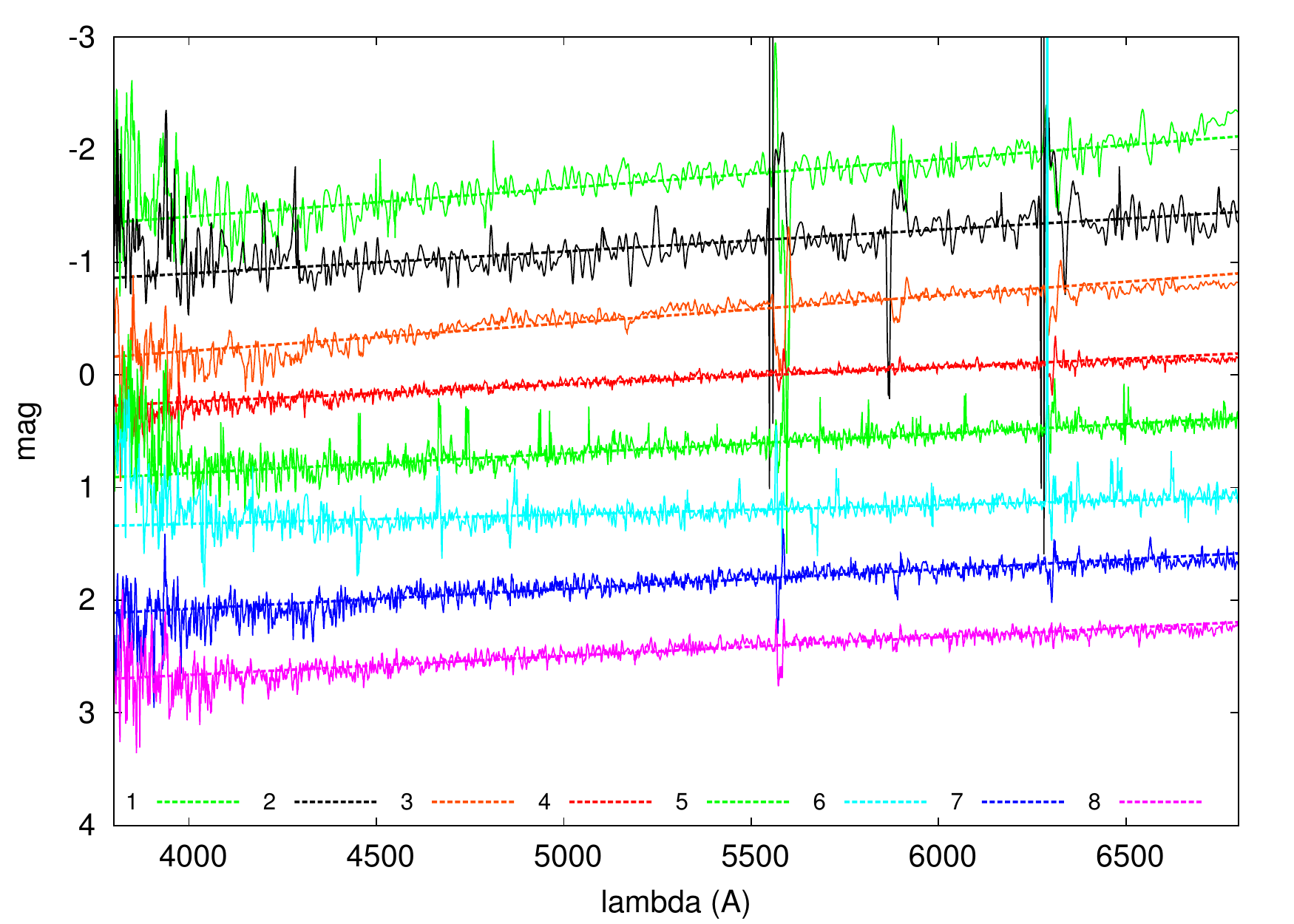}
      \caption{Average FORS spectra for each run (numbers 1-8 refer to the FORS runs in table \ref{tab:spec-obs} in chronological order), divided by Solar spectrum. The y-axis is in simple magnitudes ($-2.5\log{f}$) with arbitrary zero-points added to offset the spectra.}
         \label{avg-FORS-spec}
   \end{figure}

The $(B-V)$ index is computed from the $B$ and $V$ flux averaged over 10 nm around their respective central wavelengths; experimentation showed the result to be insensitive to the precise width of `filter' used. Corrections were made for the small difference in colour between the solar analogues used for each night and $(B-V)_\odot = 0.64$ \citep{Holmberg06}. There is scatter within each night, but the averages for each set of observations are all consistent with the overall mean value, $(B-V)$ = 0.83 (fig. \ref{BV-FORS-spec}), which is redder than the Sun but identical to the inactive nucleus colour found by \citet{Tubiana11} using broadband photometry. 
Figure \ref{avg-FORS-spec} shows the average spectra for each run, divided by the solar spectrum, showing the red slope of the continuum (dust coma). 
The ratio of the comet spectrum to the solar analogue shows a very smooth reddening function: 
The spectral slope between 440 and 540 nm is 18 \% / 100 nm, with a slight decrease in slope apparent at longer wavelengths. This is discussed further below when the wider wavelength-range X-SHOOTER data are considered. 

   \begin{figure}
   \centering
   \includegraphics[width=\columnwidth]{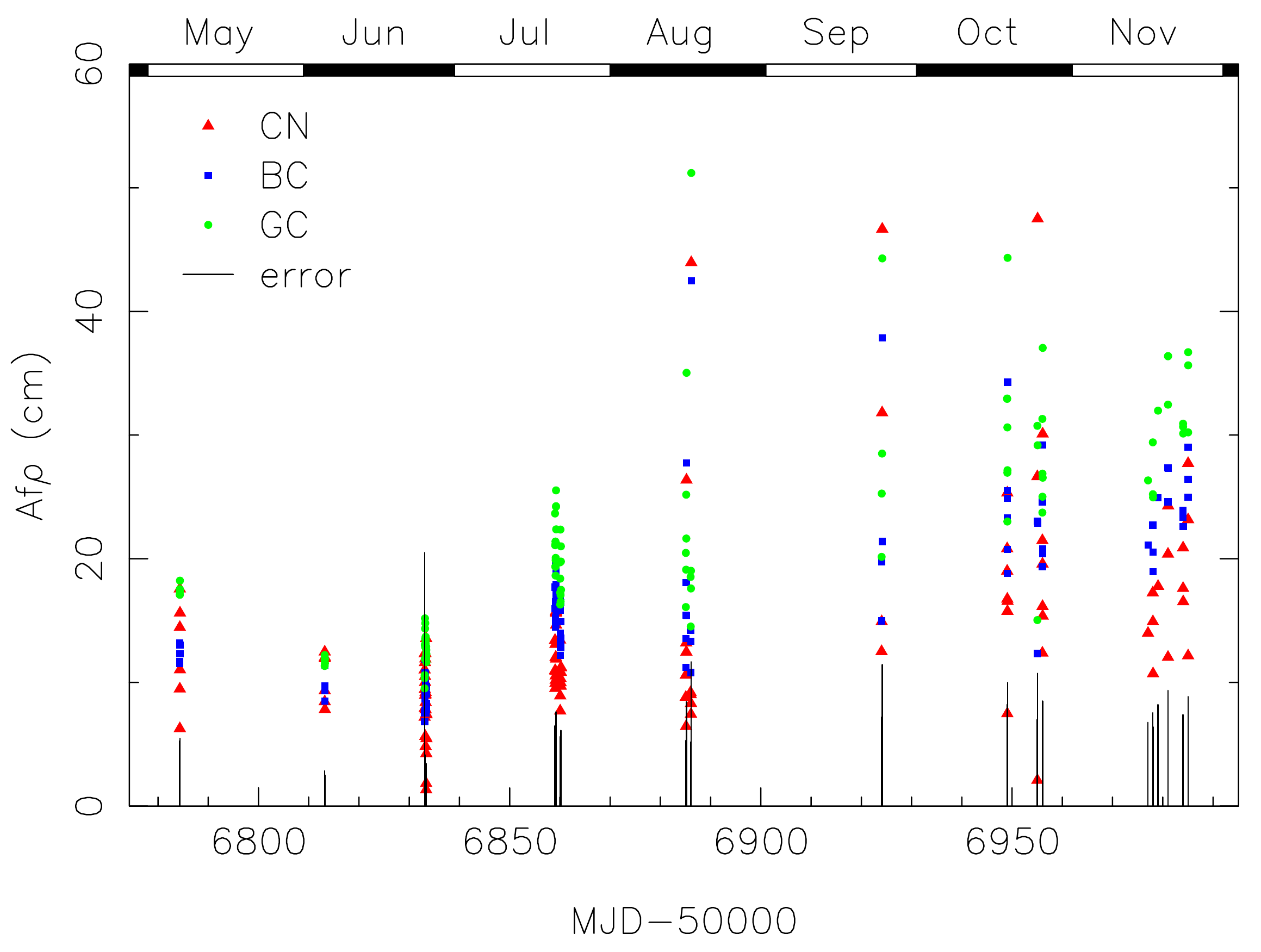}
      \caption{$Af\rho$ for all FORS spectra. The `errors' show the ranges of values that should be added if the centring was off by 0.8 arcsec. Note that the CN values are }
         \label{afrho-FORS-spec}
   \end{figure}

In the absence of emissions, a rough estimate of $Af\rho$ can be made at any wavelength. Instead of $\rho$ (the radius of the aperture, which is multiplied by $A$, the geometric albedo, and $f$, the filling factor, to give the normal $Af\rho$ quantity), we used an `effective radius' giving the area of the extraction window. 
In order to estimate the usual $Af\rho$ over a circular area, we used a model of the dust coma with a 1/$\rho$ profile convolved with a gaussian of FWHM=0.8\arcsec{}, representative of the seeing during the observations. We integrated $Af$ over the actual slit area for each spectrum and compared to the circular value. This allowed us to transform the observed  slit $Af\rho$ to the circular $Af\rho$. Moreover, we could estimate the effect of centring errors by simulating a 0.8\arcsec{} offset across the slit.    
Figure \ref{afrho-FORS-spec} shows the $Af\rho$ value measured in three wavelength regions, corresponding to the bandpasses of the CN, blue and green continuum filters of the Hale-Bopp filter set \citep{Farnham2000}, for each spectrum. The scatter of the plot is mainly due to the meteorological conditions and the centring which both lead to underestimating $Af\rho$. The average values for each run in the dust continuum filters, corrected to zero phase using the Schleicher phase function, are given in table \ref{tab:spec-results}. These values are consistent with those found through photometry using the FORS images in the previous section.
There is also a wavelength dependency of a few percent because of the reddening of 67P, apparent between the blue and green continuum, and also between these values and those determined in the red filter via photometry.

   \begin{figure}
   \centering
   \includegraphics[width=\columnwidth]{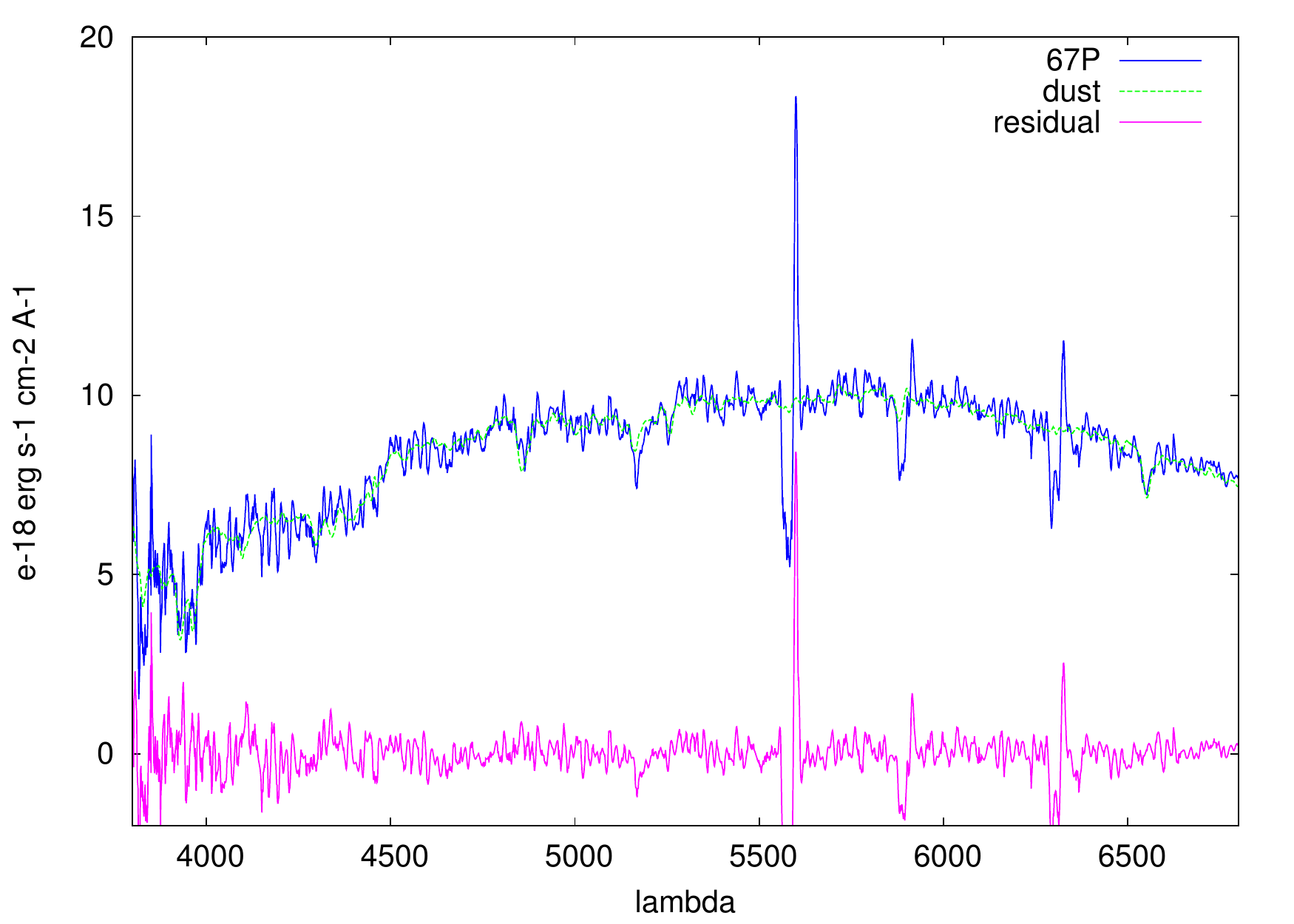}
      \caption{Example FORS spectrum (average of all individual spectra from the night of June 24th), showing the comet spectrum (blue), the scaled solar analog (green) and the residual after subtracting this (pink). The larger spikes are residual telluric lines not completely removed by the sky subtraction.}
         \label{FORS-spec}
   \end{figure}
   
      \begin{figure}
   \centering
   \includegraphics[width=\columnwidth]{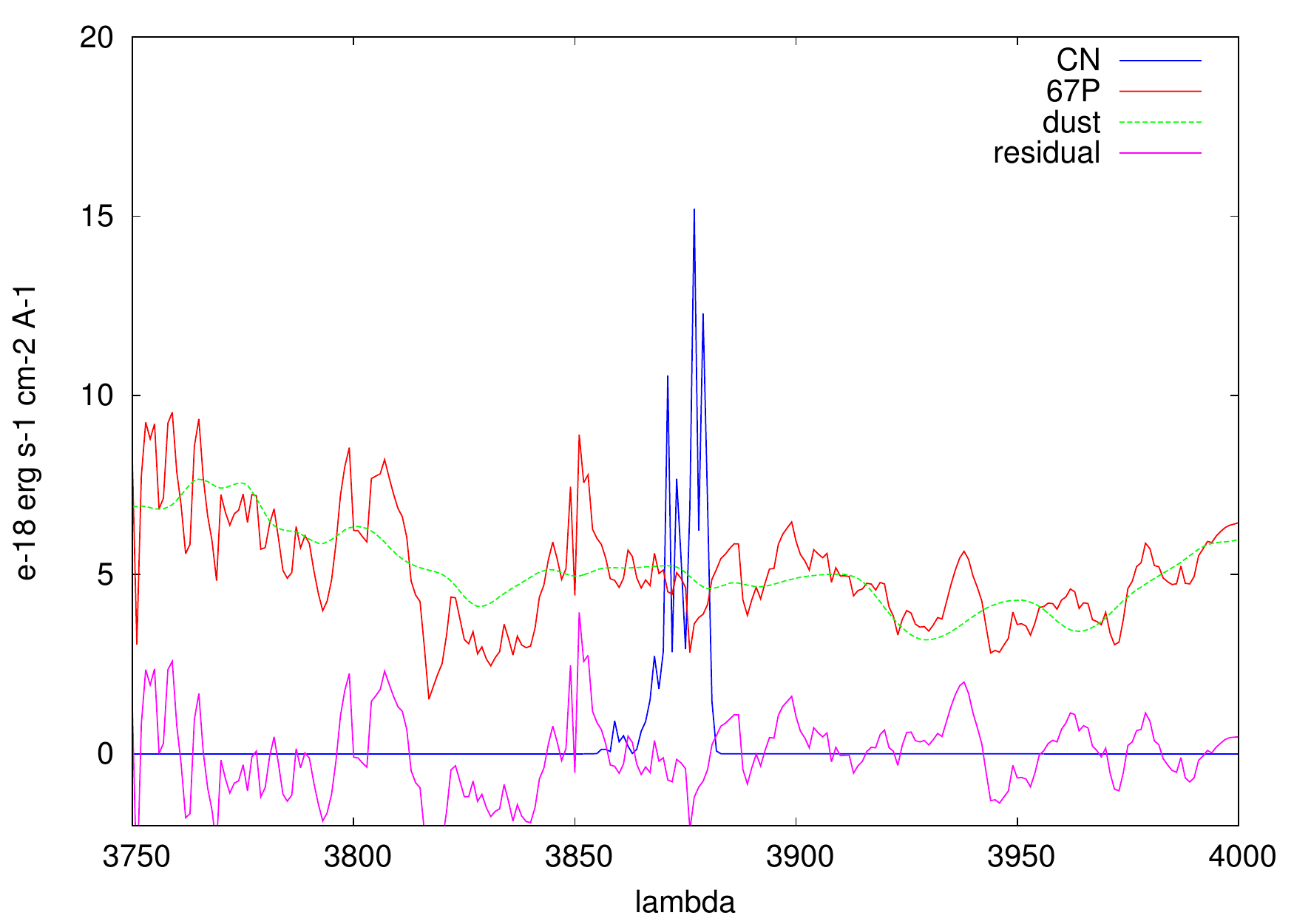}
      \caption{Zoom in on the region around the CN emission band for the FORS spectrum shown in fig.~\ref{FORS-spec}. Here the comet spectrum is shown in red, the scaled solar analog in green, and the residual after subtracting this in pink, compared with a theoretical CN emission band for Q(CN) = $5\times10^{23}$ molec. s$^{-1}$ (blue).}
         \label{FORS-spec-zoom}
   \end{figure}

The average spectrum for the June FORS2 run is shown in fig.~\ref{FORS-spec}, with a zoom in on the region around the CN band at 3875 \AA~shown in fig.~\ref{FORS-spec-zoom}. Figure~\ref{FORS-spec-zoom} also shows a synthetic spectrum of CN at a production rate of $5\times10^{23}$ molecules s$^{-1}$ for reference. CN is expected to be the strongest emission feature in the visible range. This spectrum is representative of all FORS spectra acquired, with typical S/N.
No emission can be detected in any of the spectra. Upper limits to the production of CN, C$_2$ and C$_3$ are estimated using a Haser model integrated over the slit area corresponding to the comet and the one corresponding to the background (table \ref{tab:spec-results}).
The molecular data (g-factors) come from \citet{Ahearn82} for C$_2$ and C$_3$ and \citet{Schleicher10} for CN.

   \begin{figure}
   \centering
   \includegraphics[width=\columnwidth]{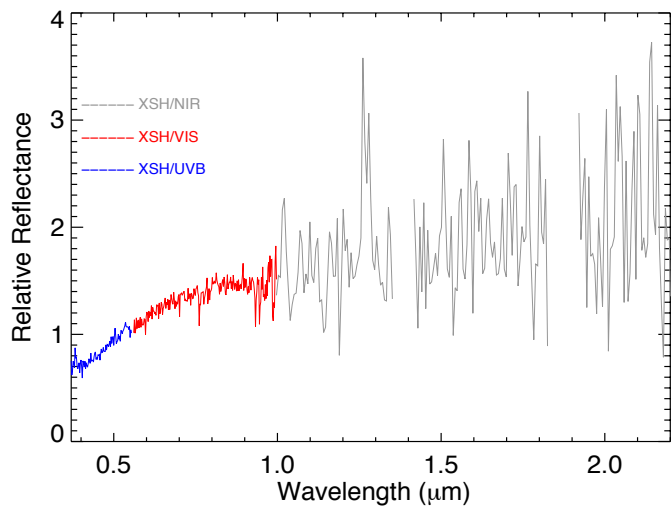}
      \caption{Average of all X-SHOOTER spectra, divided by Solar spectrum.}
         \label{XSH-spec}
   \end{figure}

   \begin{figure}
   \centering
   \includegraphics[width=\columnwidth]{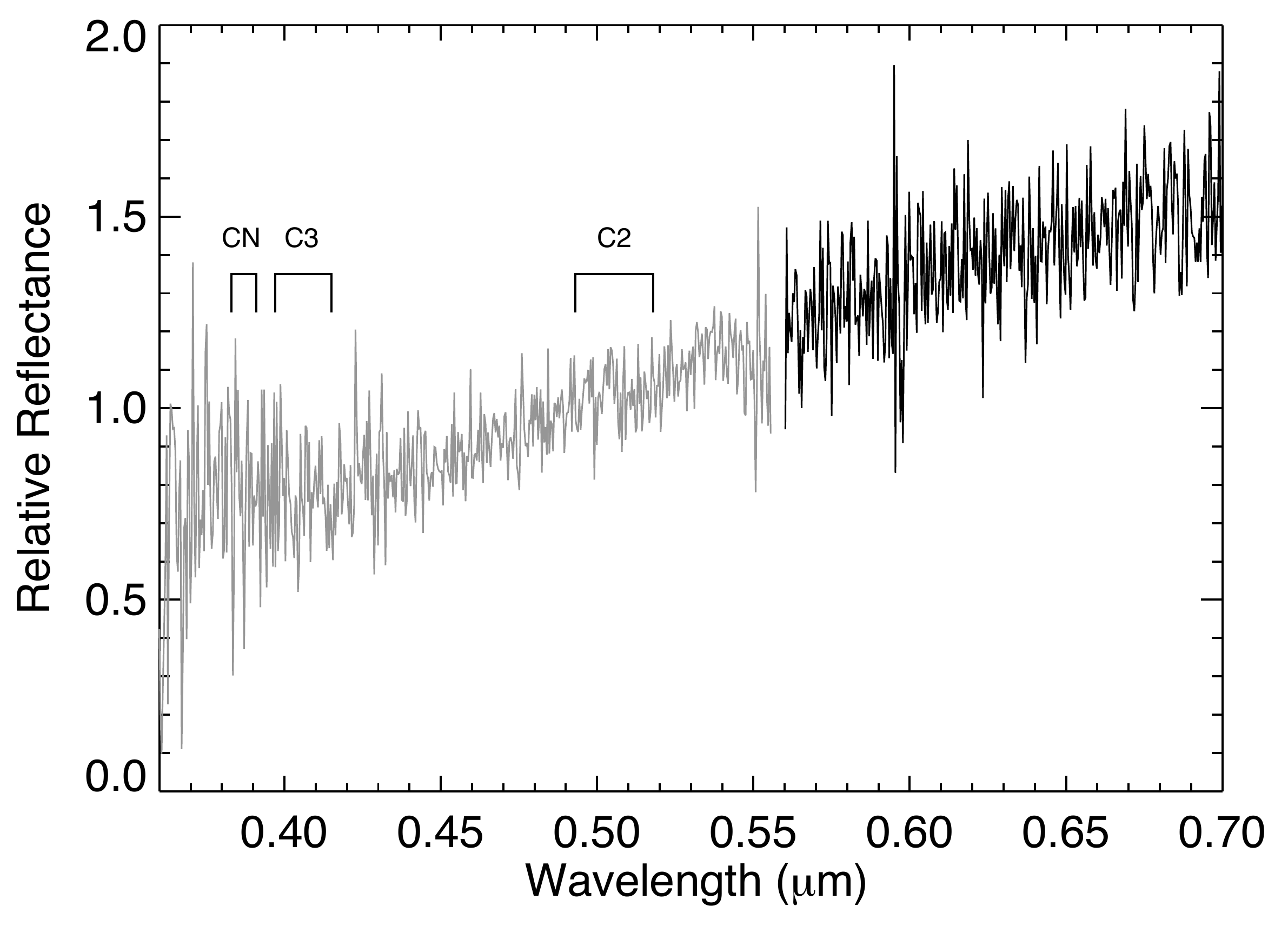}
      \caption{Zoom in on visible region of the X-SHOOTER spectrum, showing where gas emission bands would be expected.}
         \label{XSH-zoom}
   \end{figure}

The combined X-SHOOTER spectrum is shown in fig.~\ref{XSH-spec}, binned by 100 pixels in wavelength (equivalent to 2 nm in the UVB/VIS and 6 nm for $\lambda > 1000$ nm) and normalised at 550 nm, together with a zoom in on the UVB/VIS region where gas emission bands are expected, binned by 20 pixels (0.4 nm) in wavelength (fig.~\ref{XSH-zoom}). No gas emissions are detected in the X-SHOOTER spectrum either, although the better blue sensitivity and resolution means that the upper limits on the production rates of CN, C$_2$ and C$_3$ are approximately two times lower than the FORS results for later in November (table \ref{tab:spec-results}). 

The wider wavelength range of X-SHOOTER allows us to look at other aspects of the spectrum. There is no evidence for emissions at other wavelengths  -- the spectrum is too noisy in the bluest order to put constraints on the production rate of OH, despite the theoretical lower wavelength limit of the instrument being 300 nm. There is also no convincing sign of absorption by icy grains in the NIR region of the continuum, or of mineralogical signatures. The spectral slope is less steep above 800~nm, at only $\sim$ 4 \% / 100 nm, while the slope in the UVB arm is steeper than the FORS result, $\sim$ 24 \% / 100 nm. Between these extremes the slope gradually decreases with wavelength in the VIS arm, with values of $\sim$ 20 \% / 100 nm and $\sim$ 10 \% / 100 nm following the same trend seen for nucleus photometry: \citet{Tubiana11} found $\sim$ 15 \% / 100 nm and $\sim$ 9 \% / 100 nm for (V-R) and (R-I) respectively. A similar knee in the nucleus spectrum is reported by the VIRTIS instrument on Rosetta \citep{Capaccioni2015Sci}, who report slopes of 5 -- 25 \% / 100 nm below 800~nm and 1.5 -- 5 \% / 100 nm in the NIR.  The Rosetta/OSIRIS instrument found average spectral slopes of 11 -- 16 \% / 100 nm over 250--1000 nm \citep{Fornasier-AA}, and also see a decreasing slope with wavelength. The X-SHOOTER slopes are consistent with these results.

The shallow NIR slopes are also consistent with the photometry from Gemini, where average slopes in the $J$ -- $K$ (1.2 -- 2.2 micron) range are $\sim$ 2 \% / 100 nm in the October data set and near zero (i.e. solar spectrum) in November, around the same time as the X-SHOOTER spectrum. The November Gemini data set contains only $J$ and $K$ band photometry taken 4 nights apart, so the latter slope is relatively uncertain. 



%
%

\section{Discussion}\label{sec:discussion}

Our photometry and spectroscopy of the coma revealed a dust spectrum indistinguishable
from that of the global nucleus properties as observed by Rosetta.  
The fresh dust released into the coma at these distances was clearly similar to that
deposited on the surface in previous apparitions (seen in the widespread blanketing of the nucleus in the OSIRIS images -- \citealt{Thomas2015Sci}). The optical colours are similar to
those observed in other comets \citep{Hadamcik09}. Similarly, the
decrease in reflectance slope (redness) going from blue optical to near-infrared wavelengths
has been observed in other comets \citep{Jewitt+Meech86}.

The non-detection of CN emission in sensitive searches requires that the production rate of this species was lower than expected, due to either lower total activity of the comet, a globally lower abundance of CN's parent ice (thought to be HCN) in 67P or a seasonal effect meaning that less of this species was released pre-perihelion. 

We observed that the total activity of the comet matched predictions, based on the brightness of the released dust, and can also confirm activity from Rosetta observations. It is worth noting that while the total dust brightness followed the prediction from \citet{Snodgrass2013}, and showed activity already from February, OSIRIS observations showed resolved activity \citep{Tubiana2014} before it was detectable in ground-based profiles (fig.~\ref{profiles}). Similar profiles based on early OSIRIS images show activity in March and April \citep[their fig.~2]{Tubiana2014}, showing the advantage in resolution afforded by a nearby ($\sim 10^6$ km) observing platform. The fact that the FORS profile from May 3rd appears stellar shows that even relatively `high' activity as seen by OSIRIS (following a clear outburst at the end of April) can be hidden within the seeing disc. The resolved coma in OSIRIS images at this time is roughly 1000 km in diameter \citep{Tubiana2014}, corresponding to 0.4\arcsec, so we would not expect to resolve it from Earth. It is worth keeping this in mind when assessing activity levels in snapshot observations of other distant comets \citep[e.g.][]{Lowry03,Snodgrass08} and in size distribution measurements based on them \citep[e.g.][]{Snodgrass11}.

In terms of gas activity, observations from MIRO revealed total H$_2$O production rates of $\sim1$, $\sim2$ and $\sim4\times10^{25}$ molecules s$^{-1}$ at distances of $r$ = 3.9, 3.6 and 3.3 au respectively \citep{Gulkis2015Sci}, which are broadly in line with the predicted rates. A more detailed study of MIRO data taken in August ($r \sim 3.5$ au) revealed that there are variations in water production rate of a factor of two over the comet's rotation and up to a factor of 30 depending on the spatial location of the MIRO footprint on the comet \citep{Lee2015}, but these effects will average out when considering the whole coma. Together with our spectroscopy, the \citet{Gulkis2015Sci} water production rates imply lower limits on the ratio Q(H$_2$O)/Q(CN) of $ > \sim5$, $\sim30$ and $\sim40$ at the distances given above. This implies that the limits on Q(CN) found are not especially constraining, given that a `typical' comet has a H$_2$O/CN ratio of $\sim300$ \citep{Ahearn95}. 

Comparing the MIRO water production rates to our dust production rates  (approximated from $Af\rho$ measurements) gives dust/gas mass ratios of $\sim$ 30, 20 and 10, much higher than the in situ Rosetta result of $4 \pm 2$ \citep{Rotundi2015Sci}, but these should really be considered (weak) upper limits; as discussed previously, $Af\rho$ has to be treated with caution when the nucleus contributes to the flux. We can state that 67P is a relatively dusty comet, as the water production rates were in line with predictions, but the total brightness is in excess of that implied by a `standard' conversion from Q(H$_2$O) by the \citet{Jorda08} relationship (see fig.~\ref{HLC} and also the discussion by \citet{Snodgrass2013}). The high total brightness could be related to the low dust velocities measured by OSIRIS and GIADA on Rosetta \citep{Rotundi2015Sci}, implying that more dust than expected remains in the aperture. Alternatively, or in addition, it could be that the dust production of 67P is driven more by the release of other volatiles (e.g. CO, CO$_2$) than water, compared with the average comet. Finally, it could be that dust is more efficiently lifted by gas in 67P than expected. Rosetta observations reveal that the dust grains are split into two populations, fluffy and compact, with the compact ones possibly connected with active areas where gas densities are higher \citep{DellaCorte2015}. If the fluffy grain population represents an additional slow-moving component generated outside the main gas `jets', this could also explain the higher than expected dust brightness and high dust-to-gas ratio. 


   \begin{figure}
   \centering
   \includegraphics[width=\columnwidth]{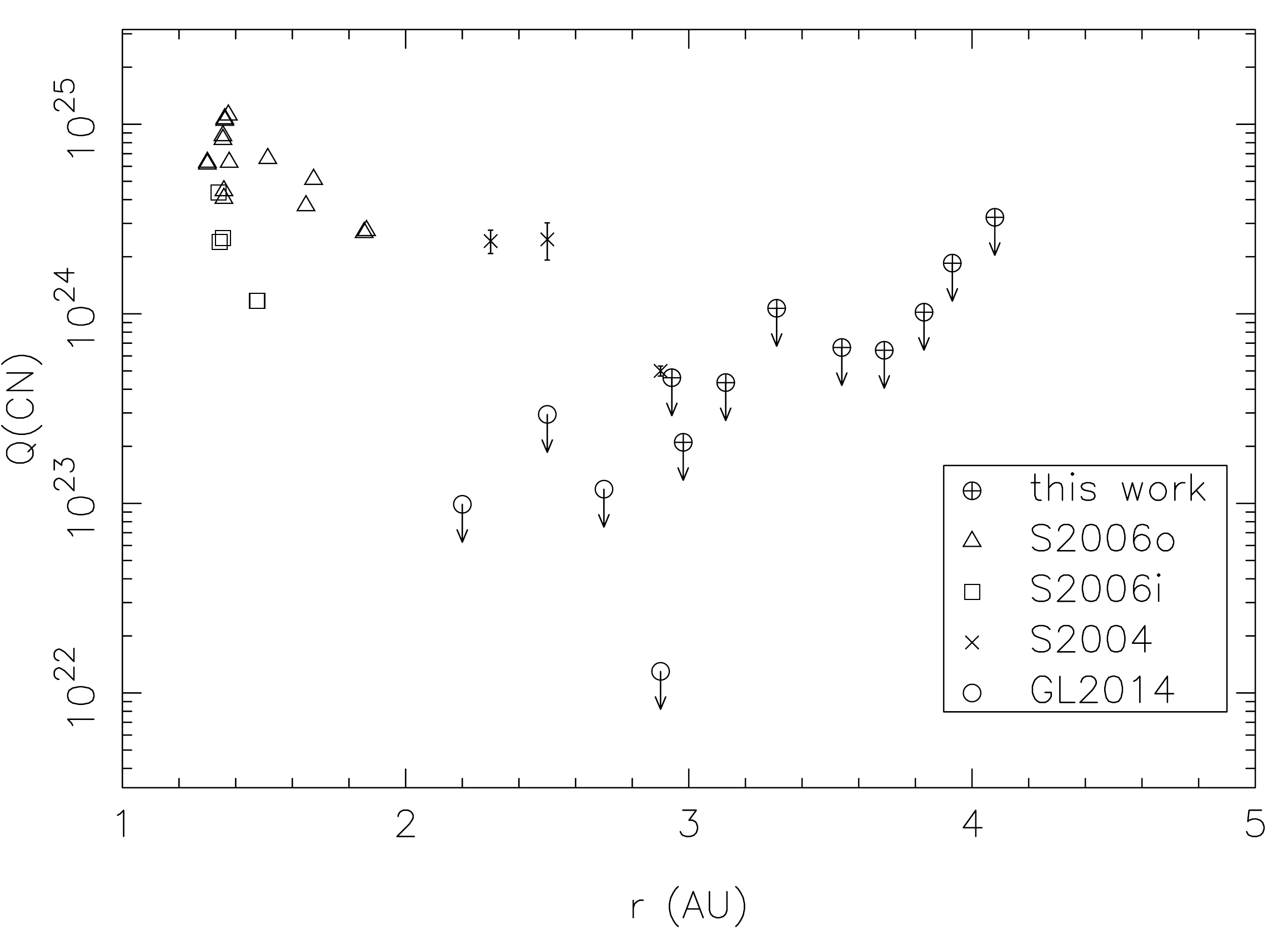}
      \caption{Production rate of CN (molecules s$^{-1}$). We include upper limits from this work (crossed circles) and \citealt{Guilbert2014} (GL2014 -- open circles), all pre-perihelion, together with post-perihelion measurements from \citealt{Schulz2004} (S2004 -- crosses). Measurements from photometry (\citealt{Schleicher2006} -- S2006) are split into inbound (squares) and outbound (triangles).}
         \label{CN-fig}
   \end{figure}

If we compare our upper limits on Q(CN) with  measurements from previous orbits (fig.~\ref{CN-fig}) we see that there is a significant asymmetry around perihelion, with clear CN detections at the same distance outbound as stronger upper limits are measured inbound. Our data, together with previous VLT/FORS  observations taken pre-perihelion in 2008 by \citet{Guilbert2014}, clearly show that we would be sensitive to production rates of order a few times $10^{23}$ molecules s$^{-1}$ at most distances, and no such activity was seen, even at close to 2 au inbound. \citet{Schulz2004}, meanwhile, detected CN at $\sim2\times10^{24}$ molecules s$^{-1}$ around 2.5 au post-perihelion in 2003, and still measured gas production close to 3 au. It does not seem likely that this change is due to significant changes in the total activity of the comet between orbits, as the total dust brightness of the comet is consistent from orbit-to-orbit \citep[see section \ref{sec:phot} and][]{Snodgrass2013}. It can also be seen that the photometry results from \citet{Schleicher2006} produce very consistent gas production rates over the 1982 and 1996 perihelion passages. These data also reveal that the asymmetry continues all the way to perihelion, with significantly lower production rates just before perihelion to the months immediately after it.

We assume that HCN is the  molecule that produces the observed CN in cometary comae. The observed lower-than-expected abundance and the asymmetry around perihelion can be explained by either 67P having generally less HCN than the average Jupiter Family comet, and we should expect orders of magnitude higher outgassing from all species after perihelion, or that there is a variation in the relative abundances with seasons on the comet.  Significant differences in the relative abundance of major species (e.g. CO vs H$_2$O) were observed in a diurnal cycle \citep{Haessig2015Sci}, but these will average out in the coma seen on ground-based observation scales.

Measurements of the HCN abundance by the ROSINA instrument on board Rosetta were made on the 19th and 20th of October 2014 \citep{LeRoy2015}, during the period covered by our FORS observations. The H$_2$O/HCN abundance ratio was observed to be very different between the illuminated and winter hemispheres at that time, at 1111 and 161 respectively. It is worth noting that the total gas density observed by Rosetta was a factor of 3.7 lower during the winter hemisphere observations, but this is not in itself enough to explain the difference (i.e. assuming that the gas pressure is dominated by water, but with absolute HCN production constant). Further ROSINA results demonstrate that HCN and water production are generally correlated -- i.e. their production rates vary in the same way with time \citep{LuspayKuti2015}.

Seasonal variation will depend on the illumination of different areas of the nucleus at different times in the orbit. We now know that the illumination of the comet around its orbit is split into a short and intense southern summer, and a much longer period when the northern hemisphere is illuminated \citep{Sierks2015Sci}. The comet experiences equinox at 1.8 au inbound and 2.5 au outbound \citep{Keller2015-spinup}. The `winter' hemisphere seen in the observations by \citet{LeRoy2015}, with higher abundance of HCN, corresponds to the southern one that is illuminated at perihelion. If more HCN is produced from the southern hemisphere this could explain the higher production rates observed by \citet{Schulz2004}, and the relatively sharp drop observed after 2.5 au outbound, but it cannot  explain the asymmetry around perihelion seen by \citet{Schleicher2006}, whose data were all taken while the same hemisphere was illuminated. Instead the more likely explanation appears to be thermal lag, where release of gas peaks some months after maximum insolation. Observations from Rosetta on either side of perihelion should allow separation of seasonal and location effects.



\section{Conclusions}

We present results from the monitoring of comet 67P using large aperture telescopes in Chile in 2014, during the pre-landing phase of the Rosetta mission. We find:

\begin{enumerate}

\item The comet was already becoming active at the start of 2014. Activity was detectable via excess flux from February (prior to the first Rosetta/OSIRIS observations), via an extended surface brightness profile from June, and directly by eye via an extended morphology from July onwards. 

\item The beginning of activity and total dust brightness provide a very good match to the predictions of \citet{Snodgrass2013}, indicating that the activity in 2014 was typical for this comet and a good match to the activity at similar heliocentric distance in previous orbits -- there does not appear to be any change in dust production from orbit to orbit for 67P.

\item No gas emission bands were detected, despite a sensitive search using both FORS and X-SHOOTER at the VLT. Upper limits demonstrate that the gas (CN) production rate is significantly lower pre-perihelion than previous observations at the same distance on the outbound leg of the comet's orbit,  with the strongest limit being Q(CN) $\le 2.1\times10^{23}$ molec.~s$^{-1}$ at Philae's landing time ($r$=3 au). This implies a high dust-to-gas ratio in this period, in agreement with Rosetta results \citep{Rotundi2015Sci}. 

\item The higher Q(CN) values appear to correspond to the period when the southern hemisphere of the comet is illuminated (i.e. the area that sees a short and intense summer around perihelion). In addition there is an indication of asymmetry in gas production close to perihelion \citep{Schleicher2006}, presumably due to a thermal lag.

\item The dust spectrum is featureless and consistent with measurements of the nucleus spectral slope, with a slope of around 20 \% / 100 nm in the $B$ -- $V$ region and a decrease in slope with wavelength, reaching $\sim$ 4  \% / 100 nm in the NIR. These slopes are consistent with the values found for the nucleus by Rosetta/VIRTIS \citep{Capaccioni2015Sci}.

\end{enumerate}

\begin{acknowledgements}

We thank the referee, Mike A'Hearn, for helpful comments that improved this manuscript.
We thank the observatory staff who made the observations possible, especially the service mode observers.
CS received funding from the European Union Seventh Framework Programme (FP7/2007-2013) under grant agreement no. 268421 and from the STFC in the form of an Ernest Rutherford Fellowship.
EJ is a FNRS Research Associate, JM is an Honorary Research Director of the FNRS and CO acknowledges the support of the FNRS. 
AF is supported by UK Science and Technology Research Council grant ST/L000709/1.
MMK is supported by NASA Planetary Astronomy Program grant NNX14AG81G. 
DMB was supported by NPRP grant X-019-1-006 from the Qatar National Research Fund.
This work was based, in part, on observations obtained at the Gemini Observatory processed using the Gemini IRAF package, which is operated by the 
Association of Universities for Research in Astronomy, Inc., under a cooperative agreement 
with the NSF on behalf of the Gemini partnership: the National Science Foundation 
(United States), the National Research Council (Canada), CONICYT (Chile), the Australian 
Research Council (Australia), Minist\'{e}rio da Ci\^{e}ncia, Tecnologia e Inova\c{c}\~{a}o 
(Brazil) and Ministerio de Ciencia, Tecnolog\'{i}a e Innovaci\'{o}n Productiva (Argentina).
This publication makes use of data products from the Two Micron All Sky Survey, which is a joint project of the University of Massachusetts and the Infrared Processing and Analysis Center/California Institute of Technology, funded by the National Aeronautics and Space Administration and the National Science Foundation.

\end{acknowledgements}

\bibliographystyle{aa}
\bibliography{comets}
\end{document}